\documentclass[aip,twocolumn,reprint,showpacs,superscriptaddress,footinbib,groupedaddress]{revtex4-1}
\usepackage{comment}
\usepackage{graphicx} 
\usepackage{dsfont} 
\usepackage{amsmath} 
\usepackage{verbatim} 
\usepackage{graphicx} 
\usepackage{ulem} 
\usepackage{units} 
\usepackage{version} 
\usepackage{color} 
\usepackage{colortbl} 
\usepackage{xcolor}  
\usepackage{fancyref} 
\usepackage[colorlinks=true,citecolor=black,urlcolor=blue,linkcolor=black]{hyperref} 
\usepackage{array} 
\usepackage{booktabs} 
\usepackage{longtable} 

\usepackage[latin1]{inputenc}
\usepackage[T1]{fontenc}

\newcommand{\ovl}{\overline}
\DeclareMathOperator*{\theo}{\Longleftrightarrow}

\begin{document} 

\title{Compact chip-scale guided cold atom gyrometers for inertial navigation: Enabling technologies and design study}
\author{Carlos L. Garrido Alzar} 
\email{carlos.garrido@obspm.fr} 

\affiliation{SYRTE, Observatoire de Paris, Universit\'e PSL, CNRS, Sorbonne Universit\'e, LNE,  61 avenue de l'Observatoire, 75014 Paris, France}
\date{\today} 

\begin{abstract} 
This work reviews the topic of rotation sensing with compact cold atom interferometers. A representative set of compact 
free-falling 
cold atom gyroscopes is considered because, in different respects, they establish a rotation-measurement reference for  
cold guided-atom technologies. The 
review first discusses enabling technologies relevant to a set of key functional building blocks of an atom chip-based compact inertial 
sensor with cold guided atoms. These functionalities concern the accurate and reproducible positioning of  
atoms to initiate a measurement cycle, the coherent momentum transfer to the atom wave packets, the suppression of 
propagation-induced decoherence due to potential roughness, the on-chip detection, and the vacuum dynamics because of its 
impact on the sensor stability, which is due to  measurement dead time. Based on the existing enabling technologies, the design of 
an atom chip gyroscope with guided atoms is formalized using a design case that treats 
design elements such as guiding, fabrication, scale factor, rotation-rate sensitivity, spectral response, 
important noise sources, and sensor stability.
\end{abstract} 

\maketitle

\section{Introduction}
Sensing, metrology, and industrial applications of cold atom interferometers drive the increasing interest 
in this topic by the physics  and engineering communities. Indeed, atom interferometry lays 
out the working principle of cold atomic clocks,~\cite{Norman,bize} 
accelerometers and gravimeters,~\cite{kasevich1,peters,peters1,abend,kang} gradiometers,~\cite{mcguirk,rosi,tino} 
gyrometers,~\cite{riehle,gustavson1,gustavson2,canuel,zhan} and 
magnetometers.~\cite{hardman} The unprecedented high measurement sensitivity and stability  made possible by these instruments have naturally triggered the development of compact and transportable atom-based quantum 
sensors.~\cite{muquans,aosense} Among 
the most relevant applications we find high-precision geophysics,~\cite{gillot} fundamental 
physics~\cite{bouchendira, Weiss, aguilera, fixler} measurements, and inertial navigation.~\cite{savoie,dutta,barret1} For inertial navigation, atom interferometers are 
particularly important because they  provide an absolute measurement of the physical quantity of interest, be it acceleration or rotation. Focusing on rotation-sensing applications, this latter fact implies that, for instance, atom gyrometers can be used to measure and preserve the 
orientation of a carrier without external references, such 
as a Global Positioning System (GPS) signal. In geophysics, a gyrometer can be used for local monitoring of the variations in Earth's rotation rate  due 
to seismic or tectonic-plate displacements.~\cite{velikoseltsev1, velikoseltsev} In the field of general relativity, for 
example, tests of the geodetic and Lense-Thirring effects~\cite{schiff,schiff1} can also be foreseen given the  long-term 
sensitivity experimentally demonstrated in the lab.~\cite{dutta1} 

Several research groups around the world launched the development of high precision compact portable atom interferometers around 
20 years ago. At Stanford,  Kasevich \textit{et al.} built a mobile atomic gravity gradiometer prototype instrument 
(MAGGPI),~\cite{kasevich2} with which they measured the Newtonian gravitational constant \(G\) and  mapped 
the  gravity gradient of Earth. Yu \textit{et al.}  at the Jet Propulsion Laboratory (JPL) worked to develop atom interferometer inertial sensors for 
gravity mapping, geodesy, fundamental physics, and planetary science.~\cite{nanyu} In Hannover, the E. Rasel's group  
developed a 
compact dual-atom interferometer for  testing  the equivalence principle~\cite{quantus, maius} and  measuring 
rotations~\cite{berg} and the fine structure constant. A collaboration between the groups of  A. Landragin at SYRTE in 
Paris and of P. Bouyer in Bordeaux developed a portable dual-species atom interferometer for  testing  the equivalence 
principle in microgravity.~\cite{ice} A portable gravimeter has also been developed at SYRTE. It participates in international comparison 
gravimetry campaigns for the SI definition of \(g\) and the redefinition of the kilogram.~\cite{jiang}

The rotation sensing capability of an atom interferometer has been demonstrated by using the Sagnac effect,~\cite{sagnac} in which two 
waves propagating in opposite directions inside a rotating interferometer of physical area \(A\)  experience a path-length 
difference and consequently a phase shift \(\Phi\) that depends on the rotation rate \(\Omega\). With the invention of 
lasers, the realization of high-sensitivity optical Sagnac gyroscopes became 
possible, such as  optical fiber 
gyroscopes~\cite{vali} and gyrolasers,~\cite{macek} which are commonly used for inertial navigation. In particular, the fiber gyroscopes 
used nowadays can reach sensitivities on the order of \(10^{-7}\)~rad\,s\(^{-1}\) over a second of integration 
time and, in the 
case of gyrolasers, this value goes down to \(10^{-8}\)~rad\,s\(^{-1}\). Following the Sagnac derivation, the minimal phase shift that 
can be measured is \(\Phi = 2 A E \Omega/(\hbar c^2)\), which indicates that \(\Phi\) does not depend on the nature 
of the propagating wave. Therefore, for the same given area \(A\), the sensitivity of an interferometer using massive particles 
of total energy \(E\) can be several orders of magnitude higher than that of an optical interferometer. This is the argument that 
triggered the development of atom interferometry, and the technology progress on atom optics allowed the realization of 
matter-wave interferometers already in the beginning of the 1990s.~\cite{mlynek,shimizu,keith}

The first demonstrations of rotation measurements were performed, as usual, with laboratory-scale devices. In 
1991, Riehle {\it et al.} used a jet of thermal calcium atoms in their gyrometer.~\cite{riehle} In this experiment, the Sagnac 
effect was observed by monitoring the displacement of interference fringes at the output of the interferometer. In 2000 
at Stanford, Gustavson {\it et al.}~\cite{gustavson1,gustavson2} developed an atom gyroscope capable of reaching a short-term 
sensitivity of 6\(\times\)10\(^{-10}\)~rad\,s\(^{-1}\)\,Hz\(^{-1/2}\), comparable to that of the best gyrolasers at that time. They 
used a Ramsey--Bord\'e~\cite{borde} symmetric interferometer with two counterpropagating thermal jets of cesium atoms. This 
configuration allowed the discrimination between rotation and acceleration effects. The first compact cold atom 
gyroscope-accelerometer was developed by the team at SYRTE.~\cite{canuel} This device used two clouds of 
cold cesium atoms in a configuration similar to the Stanford one.~\cite{gustavson1}  Table~\ref{tab:etart} presents the state-of-the-art 
in rotation technologies.
\begin{table}[tbh]
 \begin{center}
  \begin{tabular}{l||ccc}
   \bottomrule[1.5pt]
   {\bf Gyrometer} & {\bf Sensitivity} & {\bf Stability} & {\bf \(\tau_I\)} \\
   {\bf technology} & rad\,s\(^{-1}/\sqrt{{\rm Hz}}\) & rad\,s\(^{-1}\) & min \\
   \midrule[1.5pt]
   {\it Atoms} & & & \\
   SYRTE 2018 & \(3\times10^{-8}\) & \(3\times10^{-10}\) & 167 \\
   SYRTE 2015 & \(9\times10^{-8}\) & \(10^{-9}\) & 167 \\   
   \ SYRTE 2013 & \(2.6\times10^{-7}\) & \(2.5\times10^{-8}\) & 3 \\   
   \ SYRTE 2009 & \(2.4\times10^{-7}\) & \(10^{-8}\) & 30 \\   
   \ Stanford 2005 & \(8\times10^{-8}\) & \(4\times10^{-9}\) & 30 \\
   \ Stanford 2000 & \(6\times10^{-10}\) & \(2\times10^{-9}\) & 2 \\ \hline
   {\it Mechanical} & & & \\
   \ Superconducting (GP-B) & \(5.9\times10^{-7}\) & \(3.4\times10^{-13}\) & 240 \\ \hline
   {\it Optical} & & & \\
   \ Geant G-Ring laser & \(1.2\times10^{-11}\) & \(1.6\times10^{-13}\) & 300 \\ 
   \ Navigation ring laser gyro & \(
{\approx}10^{-8}\) & \(
{\approx}10^{-9}\) & -- \\
   \ Fiber iXBlue & \(3\times10^{-7}\) & \(10^{-8}\) & -- \\ \hline
   {\it Others} & & & \\
   \ RMN & \(5\times10^{-7}\) & \(4\times10^{-9}\) & -- \\
   \ He superfluid & \(8\times10^{-9}\) & -- & -- \\ 
   \ NV center  & \(10^{-5}\) & -- & -- \\ 
  \toprule[1.5pt]
  \end{tabular}
 \end{center}
\caption{State-of-the-art in rotation sensing. Here, \(\tau_I\) is 
the integration time (1~rad\,s\(^{-1}\approx 2 \times 10^{5}\) degree\,h\(^{-1}\)).}
\label{tab:etart}
\end{table}

To develop compact rotation sensors, area-enclosing magnetic guides realized with macroscopic structures 
have been 
demonstrated. For instance, by using an array of copper-tape coils in a racetrack shape, Tonyushkin and  Prentiss demonstrated a linear magnetic 
guide.~\cite{tonyushkin} In this work, an atom interferometer with an enclosed area was 
realized by splitting 
the initial cloud along the guide axis and then translating the latter in the direction perpendicular  to the splitting. With 
this method, the authors demonstrated smooth translations over centimeter-scale distances. The current in the coils was as high as 
50 A. Thus, this moving-guide configuration could be used for rotation sensing.~\cite{wu} In addition, by using a displaced linear magnetic 
guide to enclose an area, Burke and Sackett demonstrated a scalable Sagnac atom interferometer.~\cite{burke} In this 
experiment, the visibility of an interferometer realized with 3\(\times10^4\) \(^{87}\)Rb atoms was used as a measure of the 
possible attainable enclosed area. It amounted to 0.05~mm\(^2\). Yan proposed  using this solution to realize a 
guided atom gyroscope on an atom chip.~\cite{yan} To displace the guide, state-dependent microwave potentials are generated 
by a pattern of on-chip coplanar waveguides for the microwave field. Ring guiding geometries generated with macroscopic 
structures have also been proposed and realized, including the investigation of time-orbiting 
guiding potentials~\cite{gupta,reeves,arnold} and the 
realization of time-averaged 
adiabatic potentials,~\cite{gildemeister,lesa,pandey,sherlock} 
inductively coupled ring traps,~\cite{griffin}
DC storage rings,~\cite{arnold1} and RF-dressed quadrupole ring traps.~\cite{morizot1,heathcote,chakraborty}

This paper  reviews  the various enabling technologies for the realization of compact portable cold atom gyrometers 
based on atom chips. We  mainly focus on the research achievements relevant to 
inertial navigation applications. Particular attention is devoted to atom chips because of their high potential in the implementation of compact quantum 
sensors. In Sec.~\ref{sec:AI} we review the general physical principles of atom interferometry. We then present the main 
experimental realizations of compact cold atom gyrometers using free-falling atoms. These are reference examples of key 
technological solutions to specific inertial navigation questions to be considered when designing and implementing guided atom 
interferometers. Section~\ref{sec:GAI} is devoted to guided atom interferometry and discusses the main inertial navigation 
specifications of such a configuration.  Section~\ref{sec:chips}  
overviews the different enabling 
technologies for  atom-chip-based inertial sensors. The relevant results for a sensor using guided 
atoms are presented in this section.  Section~\ref{sec:perf} discusses the relevant systematic effects and noise sources 
based on an example of a guided atom interferometer. Considering the expected sensor performance, a 
particular application to a fundamental physics experiment is presented at the end of Sec. \ref{sec:perf}. 

\section{Atom interferometers for rotation sensing}\label{sec:AI}
\subsection{Elements of atom interferometry}
In an optical interferometer, beam splitters and mirrors are used to modify the propagation mode of 
 light injected at the input port of the device. In a complementary way, laser beams modify the propagation state 
of atoms in an atom interferometer (AI). At the input of the interferometer we prepare a 
well-defined propagation mode in terms of momentum and 
internal energy of the atomic state. Atom interferometers are typically implemented with stimulated Raman 
transitions to manipulate the atomic state. They are generated by counterpropagating laser beams, following the proposal of  Bord\'e.~\cite{borde}

\begin{figure}[h]
 \begin{center}
 \includegraphics[width=3.75cm]{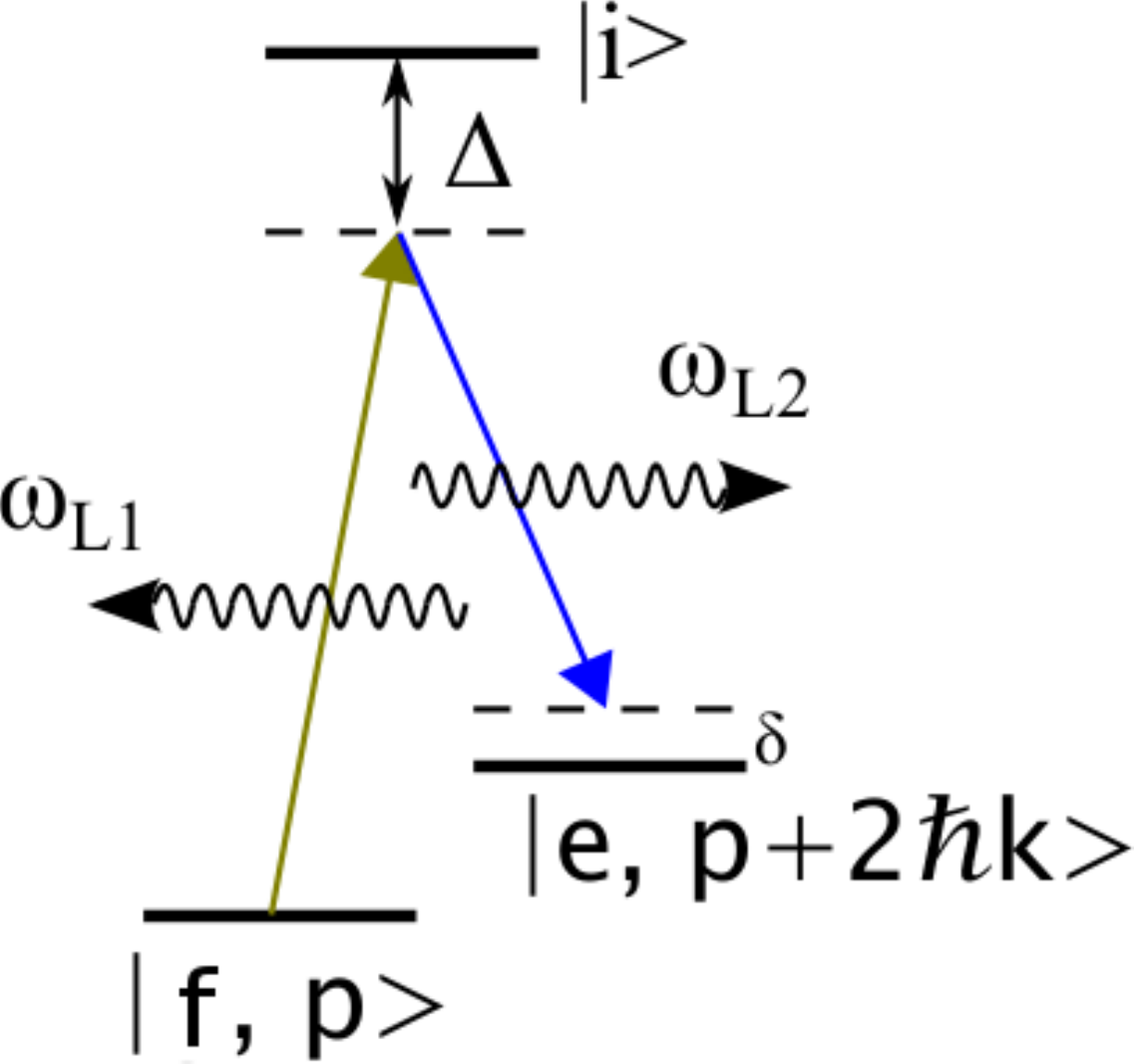}
 \caption{Representation of  stimulated Raman transitions induced by two laser beams of frequencies $\omega_{\rm L1}$ and 
 $\omega_{\rm L2}$. The Raman detuning is given by $\delta.$}
 \label{fig:transR}
 \end{center}
\end{figure}

The Raman transition is a two-photon process in which the two laser beams interact with the atoms described by a three-level 
system, as shown in Fig.~\ref{fig:transR}. Here, the laser frequencies are chosen to be significantly detuned by  \(\Delta\) from any possible optical transition involving an excited intermediate state \(|i\rangle\). In this way, 
 spontaneous emission is suppressed, thereby preserving the coherence properties of the atoms. 

Let us consider a particular realization of the process sketched in Fig.~\ref{fig:transR}. Suppose that the atom is initially 
prepared in an internal state of energy \(|f\rangle\) and  momentum \(
{\bf p}\). After the absorption of a photon at  
frequency \(\omega_{\rm L1}\) and the subsequent stimulated emission of a photon at \(\omega_{\rm L2}\), the  internal state of the atom 
is \(|e\rangle\) and its momentum \(
{\bf p}+\hbar ({\bf k}_{\rm L1} - {\bf k}_{\rm L2})\). Because the laser beams propagate 
in opposite directions, the atom  changes its initial momentum by \(2\hbar k\), where 
\(2k=2|{\bf k}|=|{\bf k}_{\rm L1} - {\bf k}_{\rm L2}|\). To estimate the order of magnitude of the corresponding change in velocity, we 
consider a transition on the \(D_2\) line of \(^{87}\)Rb atoms. Along the axis of the Raman beams, the change in velocity  is
 \(2\times5.9\)~mm/s, which is twice the recoil velocity. This implies that, if the atom's state changes from \(|f, {\bf p}\rangle\) to 
\(|e, {\bf p}+2\hbar {\bf k}\rangle\), then after one second its initial trajectory will be deflected by 1.2~cm.

\begin{figure}[h]
 \begin{center}
 \includegraphics[width=\columnwidth]{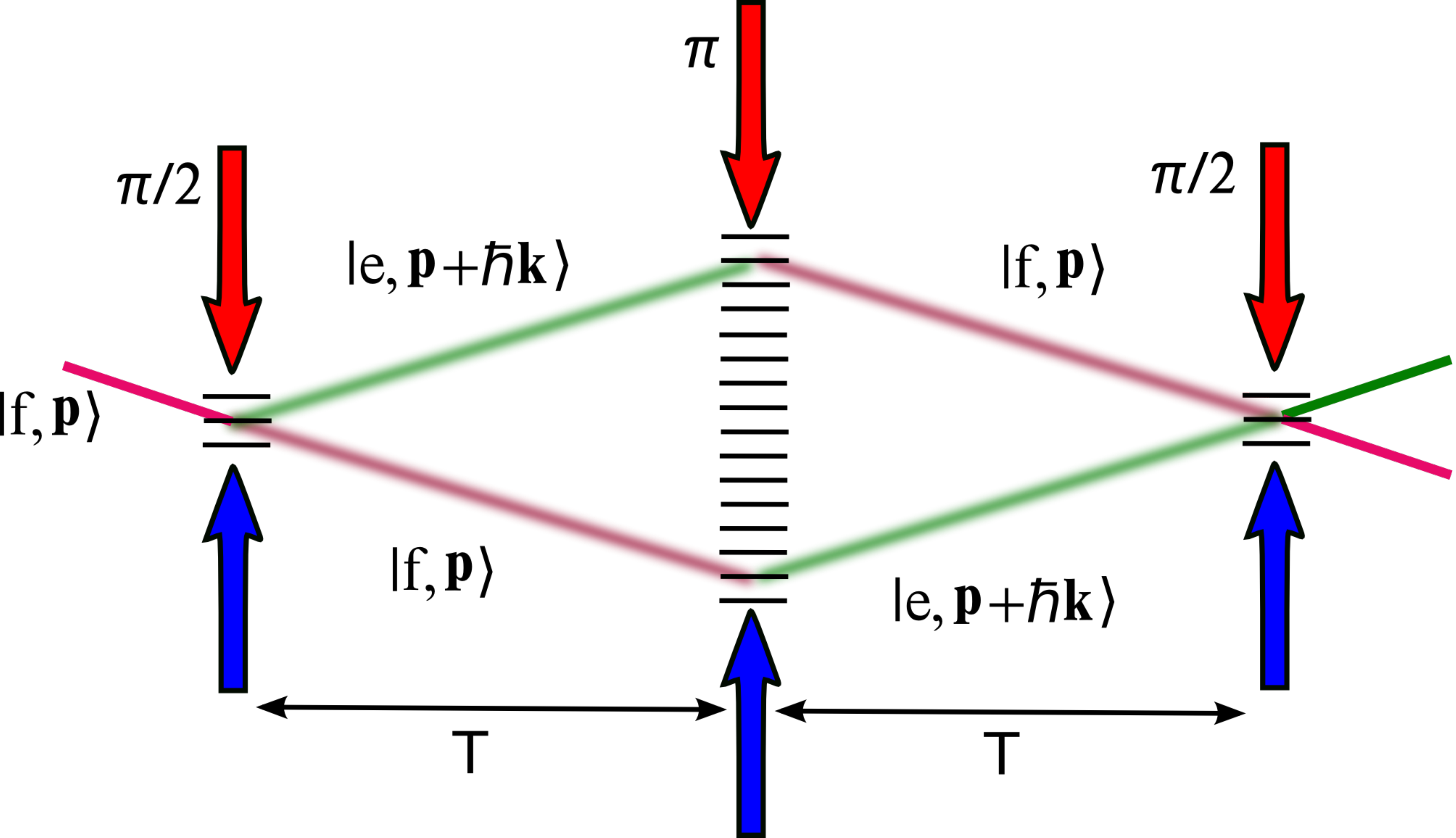}
 \caption{Mach--Zehnder atom interferometer configuration. The  short, black, horizontal lines represent the equiphase planes 
 of the laser beams in the laboratory frame, and \(T\) is the free propagation time.}
 \label{fig:ramanMZ}
 \end{center}
\end{figure}

By using light pulses (i.e., by changing the duration of the interaction between the atoms and the laser beams),  any superposition state may be prepared between \(|f, {\bf p}\rangle\) and \(|e, {\bf p}+2\hbar {\bf k}\rangle\). In particular, the 
pulse duration that prepares a superposition of the states with the same weight for both states is called a ``\(\pi/2\)'' 
pulse and is used to implement a matter-wave beam splitter. If the light pulse exchanges the two states, then it is a  ``\(\pi\)'' pulse and it implements a matter-wave mirror. Typically, these pulses have a duration of a few tens of 
microseconds for Raman beams with an optical power of about 400 mW and a 1/\(e^2\) radius of 2~cm. The standard geometry of an 
atom interferometer is of Mach--Zehnder type with three light pulses, as shown in Fig.~\ref{fig:ramanMZ}.

To determine the atomic phase \(\Phi\) at the output of the AI, we measure the probability of detecting the atoms in  states 
\(|f, {\bf p}\rangle\) and \(|e, {\bf p}+2\hbar {\bf k}\rangle\), which   can be written as 
\(P=A+B \cos(\Phi)\). For an apparatus subjected to an acceleration {\bf a}, the phase \(\Phi\) can be shown to be given by~\cite{borde}
\begin{equation}
 \Phi_{\rm a}= {\bf k}\cdot {\bf a}\ T^2,
\label{eq:phiacc0}
\end{equation}
where \(T\) is free propagation time between the laser pulses. If the device is subjected to a rotation at an 
angular rate \(
{\bf \Omega}\), then the atomic phase is
\begin{equation}
 \Phi_\Omega= 2{\bf k}\cdot ({\bf \Omega}\times {\bf v})\ T^2,
\label{eq:phirot}
\end{equation}
where \(
{\bf v}\) is the initial velocity of the atoms in the state \(|f, {\bf p}\rangle\) at the AI input. The global 
inertial phase measured by an AI is therefore \(\Phi=\Phi_{\rm a} + \Phi_\Omega\), which thus  contains information about both 
 the rotations and accelerations to which the device is subjected. This phase is imprinted by the laser beams onto the atomic 
wave function during the light-atom interactions that occur in the AI. Several techniques are available to 
isolate accelerations from rotations,  the most common being to use a dual atom cloud source or a pulse configuration that renders the interferometer 
sensitive to rotations but not to DC accelerations. For instance, if a dual cloud source is used, then the two clouds are 
launched following reciprocal paths at exactly the same absolute velocity. In this situation, at the end of the interferometer 
sequence, the global inertial phase measured on one cloud  carries a rotation contribution with  velocity \(
{\bf v}\), whereas 
the measurement of the global phase on the other cloud gives a rotation contribution with  velocity \(-{\bf v}\). By 
adding and subtracting these two global phases, one may distinguish accelerations from rotations. How 
the global phase \(\Phi\) is related to inertial forces is presented below.

\subsection{Free-falling atom compact rotation sensors}
Compact atom interferometer inertial sensors can be divided into three large classes: free-falling, 
confined, and guided 
atom interferometers. Out of these three classes, free-falling atoms define the 
state of the art, so we  consider here 
a few representative implementations of this class. Almost  all  realizations of inertial sensing with AIs use 
free-falling 
cold atoms. The atoms are either dropped, 
launched vertically in a fountain configuration, or follow parabolic trajectories. In this way, the atoms 
provide an inertial frame with respect to which the forces exerted on the instrument can be measured. For example, in 
Fig.~\ref{fig:ramanMZ} the AI is sensitive to an acceleration in the direction of the Raman beams. On the one hand, 
accelerations  translate into a displacement of the lasers' equiphase planes (short black lines) with respect to the free-fall atomic trajectories. On the other hand, rotations of the equiphase planes with respect to the atomic trajectory and 
about an axis perpendicular to the AI's oriented area are mapped to different light-atom interaction strengths in the phase of the 
atomic wave function at every light pulse.

\begin{figure}[h]
 \begin{center}
 \includegraphics[width=\columnwidth]{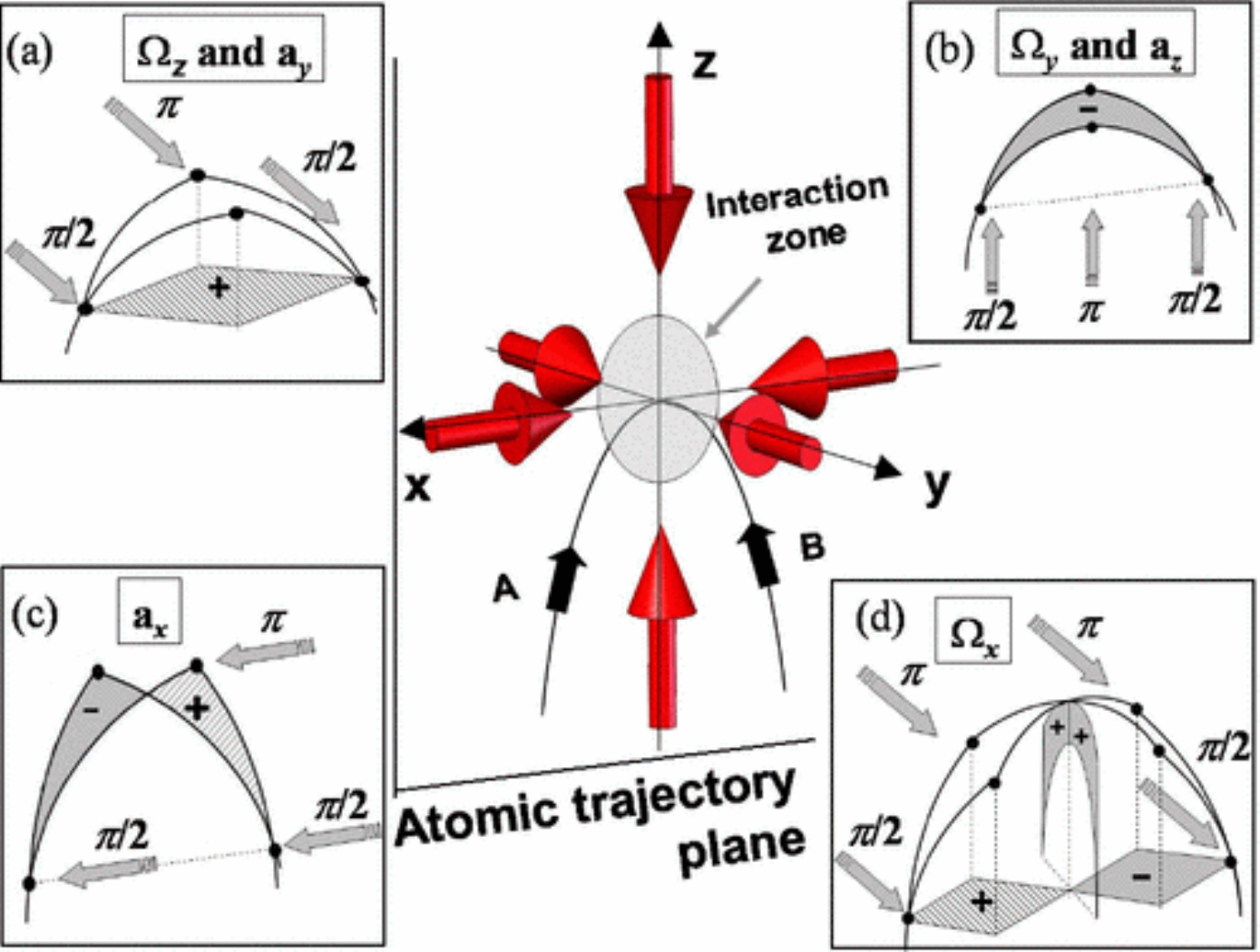}
 \caption{Principle of a six-axis inertial sensor. The atomic clouds are launched on a parabolic trajectory and interact with the 
 Raman lasers at the top. The four configurations (a)--(d) give access to the three rotations and the three accelerations. In the three-pulse configuration, the Raman beam  can be horizontal or 
 vertical. Therefore, the atom cloud can be split or deflected in (a)\ 
 the horizontal \(x\text{-}y\) plane 
  or (b), (c) in the vertical \(x\text{-}z\) plane. (d) With a butterfly four-pulse sequence of horizontal beams, the rotation \(\Omega_x\) can be 
 measured. Adapted from Canuel {\it et al.,}~\cite{canuel} \copyright\ 2006 by the American Physical Society.}
 \label{fig:gyrSyrte1}
 \end{center}
\end{figure}

\begin{figure}[h]
 \begin{center}
 \includegraphics[width=0.8\columnwidth]{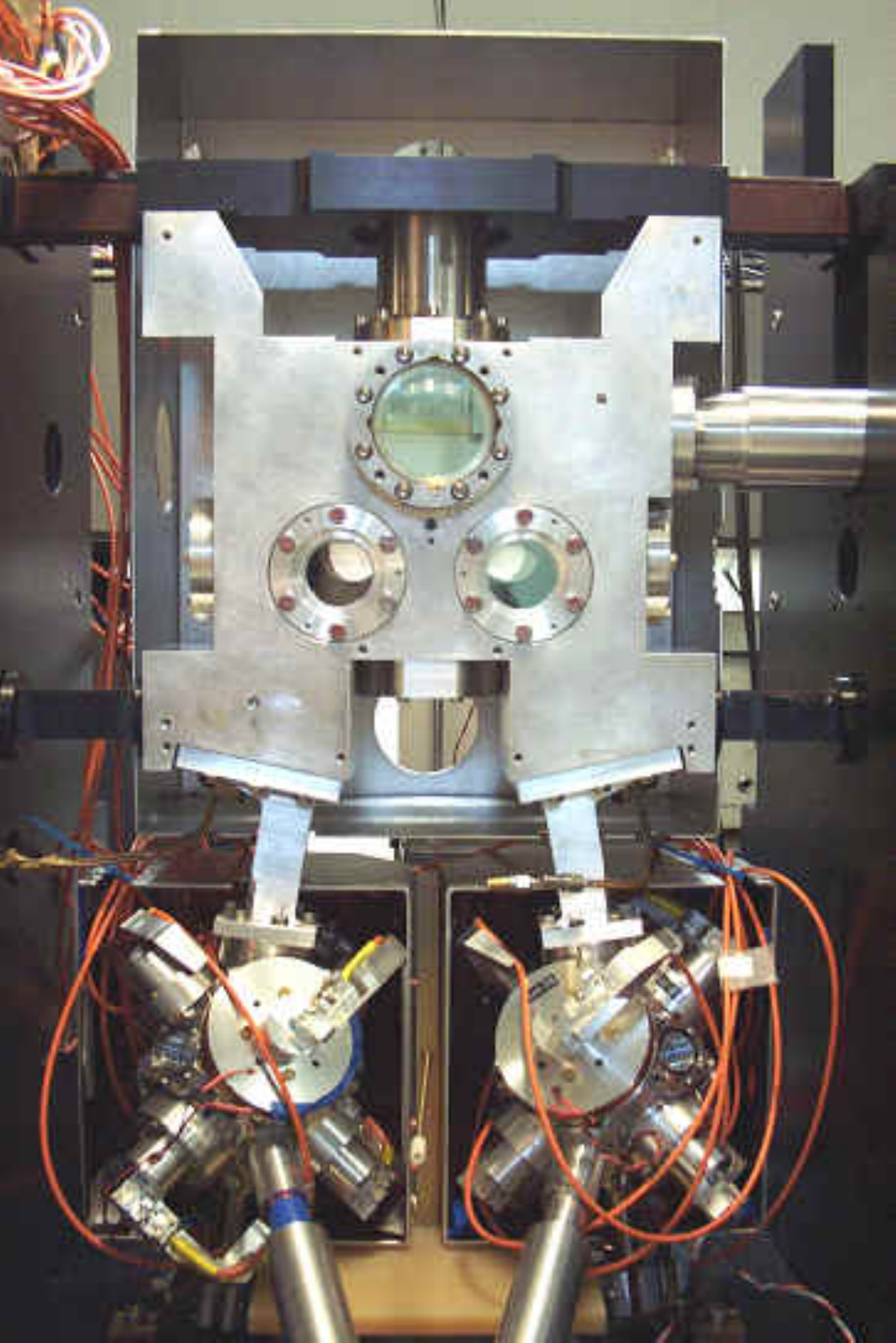}
 \caption{Picture of the SYRTE six-axis inertial sensor. The  instrument is 50~cm tall. Courtesy of A. Landragin
(Paris).}
 \label{fig:gyrSyrte2}
 \end{center}
\end{figure}

One representative realization of a compact, free-falling AI was the first cold atom gyrometer realized by Canuel 
{\it et al.} at SYRTE.~\cite{canuel} This atom interferometer used a dual cloud source and  measured the six inertial axes following the working principle presented in Fig.~\ref{fig:gyrSyrte1}. It 
offered  projection-noise-limited performance, with a sensitivity to rotations and accelerations of 
2.4\(\times\)10\(^{-7}\)~rad\,s\(^{-1}\)\,Hz\(^{-1/2}\) and 5.5\(\times\)10\(^{-7}\)~m\,s\(^{-2}\)\,Hz\(^{-1/2}\), respectively. Despite this 
remarkable result, the performance of this device was limited by the initial temperature of the atoms (2~\(\mu\)K), the 
relatively short interrogation time (\(2T= 60\)~ms), the small area (\(\approx\)4~mm\(^2\)) of the interferometer, and the superposition of the 
atomic trajectories (Fig.~\ref{fig:gyrSyrte2}).

\begin{figure}[h]
 \begin{center}
 \includegraphics[width=0.7\columnwidth]{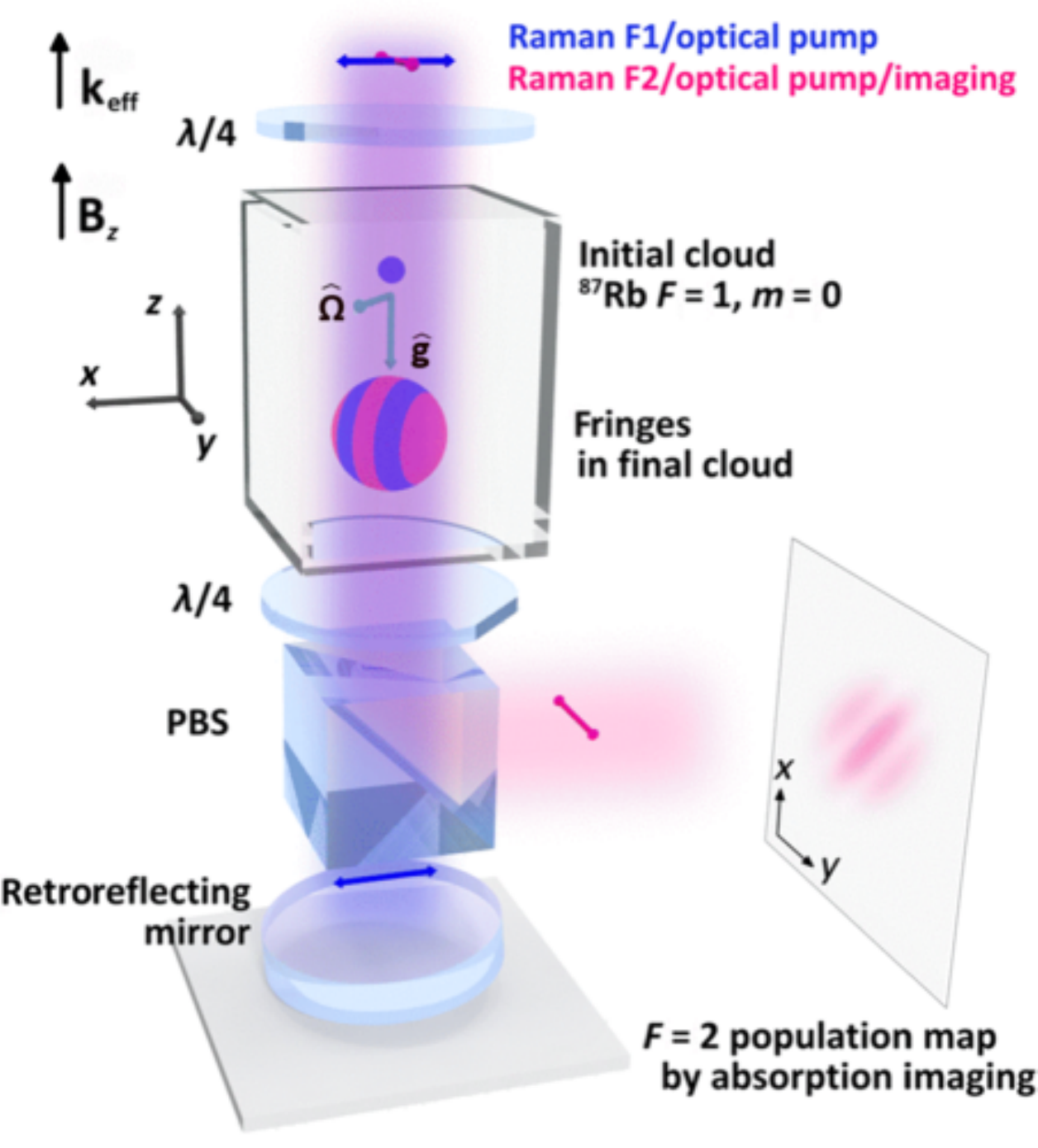}
 \caption{Schematic of the PSI science package. The inner dimension of the glass cell is 
 1~cm\(^2\). The instrument measures  accelerations in the \(z\) direction and the projection of the rotation vector onto the 
 \(x\text{-}y\) plane. ``PBS'' is a  polarizing beam splitter. Adapted from Chen {\it et al.,}~\cite{chen} \copyright\ 2019 by 
 the American Physical Society.}
 \label{fig:nist}
 \end{center}
\end{figure}

At NIST, Donley's team recently 
demonstrated a multiaxis cold-atom gyrometer by using a single source in a centimeter-scale cell.~\cite{chen} Their device  simultaneously measures accelerations and rotations by using a point source atom interferometer (PSI), as sketched in 
Fig.~\ref{fig:nist}. After  averaging 
for 1~s, the measured sensitivities for the magnitude of the rotation rate  and the direction were  \(0.033^{\circ}\)\,s\(^{-1}\) and  \(0.27^{\circ}\), respectively. Concerning acceleration measurements, the authors demonstrated a relative precision in the 
gravity measurement of \(\delta g/g = 1.6\times 10^{-6}\)~Hz\(^{-1/2}\). The PSI technique consists of an ensemble of
 single-atom independent interferometers. 
Because 
of the initial thermal velocity distribution of the cold atom cloud,  a strong correlation exists between position and 
velocity after a ballistic expansion during the free-fall time of the cloud. This correlation means that two 
different atoms with exactly opposite velocities will interact with the Raman beams in the same way as the two clouds in 
the SYRTE gyrometer. It is this correlation that is exploited to 
measure distinguishably accelerations and rotations. Consequently, the PSI can be seen as 
a collection of dual-source atom interferometers, but at the single-atom 
level. The strength of the PSI  
is that it can measure rotation and 
acceleration in a single measurement shot and using a single 
atom cloud.

Typically, the measurement rate of an atom interferometer is on the order of a Hz. However, applications 
such as inertial navigation and guidance require a high measurement bandwidth for the proper real-time integration of the 
vehicle equations of motion. At Sandia National Laboratories, Rakholia {\it et al.}~\cite{rakholia} tackled 
this critical question and  realized a dual-axis device 
with a 60 Hz bandwidth. The core idea of their technique is based on  combining a dual atomic source with the recapture 
of a fraction of the atoms after the interferometer sequence. By  simultaneously measuring 
accelerations and rotations, this work presents a building block of a six-axis inertial sensor.  Although the interferometer 
duration is relatively short (4~ms), the loss in sensitivity  is compensated by using large 
atom numbers, large launching velocities, and reducing the dead time. The measured sensitivities, which are typical of laboratory 
experiments, are in the range of \(\mu g\)\,Hz\(^{-1/2}\) and \(\mu\)rad\,s\(^{-1}\)\,Hz\(^{-1/2}\) for acceleration and 
rotation, respectively. Such performance is the result of a carefully developed device, analyzed following a model developed by the authors. The 
identification and characterization of the various noise sources  also contributed to achieving the high 
sensitivities shown in this work. 

Another example of 
free-falling compact atom gyrometers was developed at Hannover in Rasel's group.~\cite{tackmann} Their gyrometer had a Sagnac area 
of 19~mm\(^2\) and  was 13.7~cm long. The rotation measurement was performed by using a cold atom beam at 
2.79~m\,s\(^{-1}\) and 
 achieved a sensitivity of 6.1\(\times\)10\(^{-7}\)~rad\,s\(^{-1}\)\,Hz\(^{-1/2}\). 

At Bordeaux, Barrett {\it et al.} developed a dual-species \(^{87}\)Rb-\(^{39}\)K compact inertial sensor for the study of the equivalence 
principle in 
microgravity.~\cite{barret} One of the important achievements of this experiment, from the inertial-sensing perspective, is 
the operation of an atom interferometer in the strongly vibrating environment of an aircraft. The sensor, subjected not 
only to vibration levels of 10\(^{-2}\)~\(g\)\,Hz\(^{-1/2}\) but also to rotation rates as large as  \(5^\circ\)\,s\(^{-1}\),  measured the E\"otv\"os ratio in microgravity with a systematic uncertainty of 3\(\times\)10\(^{-4}\).

\section{Guided atom interferometry with atom chips}\label{sec:GAI}

\subsection{Main requirements for navigation}
For an inertial navigation application, the main constraints for an atom interferometer  are small volume, 
low power consumption, high bandwidth, high dynamic range, low cost, robustness, and device survivability when 
exposed to or 
operated in a severe environment.~\cite{barbour} The sensor is then suitable for a specific inertial navigation 
task depending on its performance. For instance, the angular random walk of a navigation-grade gyrometer has to be less than 
\(10^{-3}{}^\circ\, \text{h}^{-1/2}\).

To use an atom chip as a sensor platform, several technological obstacles have to be addressed. Whether trapped or guided, 
coherent splitting and recombination of a sub-Doppler--cooled thermal or degenerate atomic ensemble is required. If we use 
a trapped ensemble, then  symmetric splitting at the level of 10\(^{-3}\)  is required to have coherence times 
greater than 40~ms, and acceleration must be sensed at the 10\(^{-6}~g\) level. In the case of a guided atom 
interferometer, we need to fabricate or somehow generate equivalent roughness-free matter-wave guides. For rotation sensing, 
the phase bias stability has to be at the 10\(^{-4}{}^\circ \,{\rm h}^{-1}\) level.

\subsection{Enabling technologies}\label{sec:chips}
Atom chip technology has been extensively covered in several reviews.~\cite{keil,fortagh,folman,folman1,lev,reichel} Here, we  address the atom chip enabling technologies relevant 
to key functional building blocks necessary for implementing  compact inertial sensors with cold guided atoms. These 
functions concern
\begin{itemize} 
 \item the precise and accurate positioning of the atoms to start the interferometer sequence;
 \item the coherent momentum transfer and splitting of the atom clouds;
 \item the attenuation or suppression of propagation-induced decoherence;
 \item the on-chip detection;
 \item the vacuum dynamics because of its non-negligible contribution to the Dick effect~\cite{dick} via  sensor dead time.
\end{itemize}
Addressing these points is important in order to exploit the full potential for compactness offered by an atom chips.

Atom chips are a promising technology for 
 manipulating  cold atoms using complicated confining geometries, which is important for developing compact matter-wave interferometers.~\cite{keil} In fact, it is possible to microfabricate on an atom 
chip a complex wire pattern to create sub-micron magnetic potentials with the shape required by the targeted application. We 
can, for example, design arrays of potential wells for quantum information processing,~\cite{sinuco} traps for measuring accelerations,~\cite{ammar} and toroidal waveguides for detecting rotations.~\cite{clga,lesanovsky,fernholz,seungjin} 

The coherent beam splitting of a cold thermal cloud has already been demonstrated for propagating atoms.~\cite{cassetari} On 
atom chips, the coherent manipulation of Bose--Einstein-condensed trapped and propagating atoms has been observed in atom 
interferometers~\cite{schumm,maussang,shin,wang} and atomic clocks.~\cite{treutlin,deutsch}

Although cold atom propagation in circular macroscopic guides has already been observed,~\cite{gupta,gildemeister}   the demonstrated sensitivity to rotations  was insufficient  for high-precision measurements or inertial 
sensing applications. Using a linear waveguide, Wu {\it et al.} estimated the expected rotation sensitivity 
of their enclosed-area interferometer.~\cite{wu} Although the estimated
short-term sensitivity was  1\(\times\)10\(^{-9}\)~rad\,s\(^{-1}\)\,Hz\(^{-1/2}\), the expected 
stability was insufficient for measurements at a metrologically 
relevant level. In other work, Qi {\it et al.}~\cite{qi} built  a magnetically guided atom interferometer that uses Cs atoms and for which  they used a compact architecture to measure 
accelerations. The guide was generated by race-track coils carrying a current of about 30 A. The resulting 
magnetic guide had a radial frequency of 98 Hz and  was loaded with 5\(\times 10^7\) Cs atoms. The resulting measurement 
uncertainty was  7\(\times10^{-5}\)~m\,s\(^{-2}\) with an interrogation time of \(2T\) = 18~ms. The  guide had an enclosed area 
of 1.8\(\times10^{-2}\)~mm\(^{2}\) and  can potentially be used to sense rotations. 

\begin{figure}[htb]
 \begin{center}
 \includegraphics[width=\columnwidth]{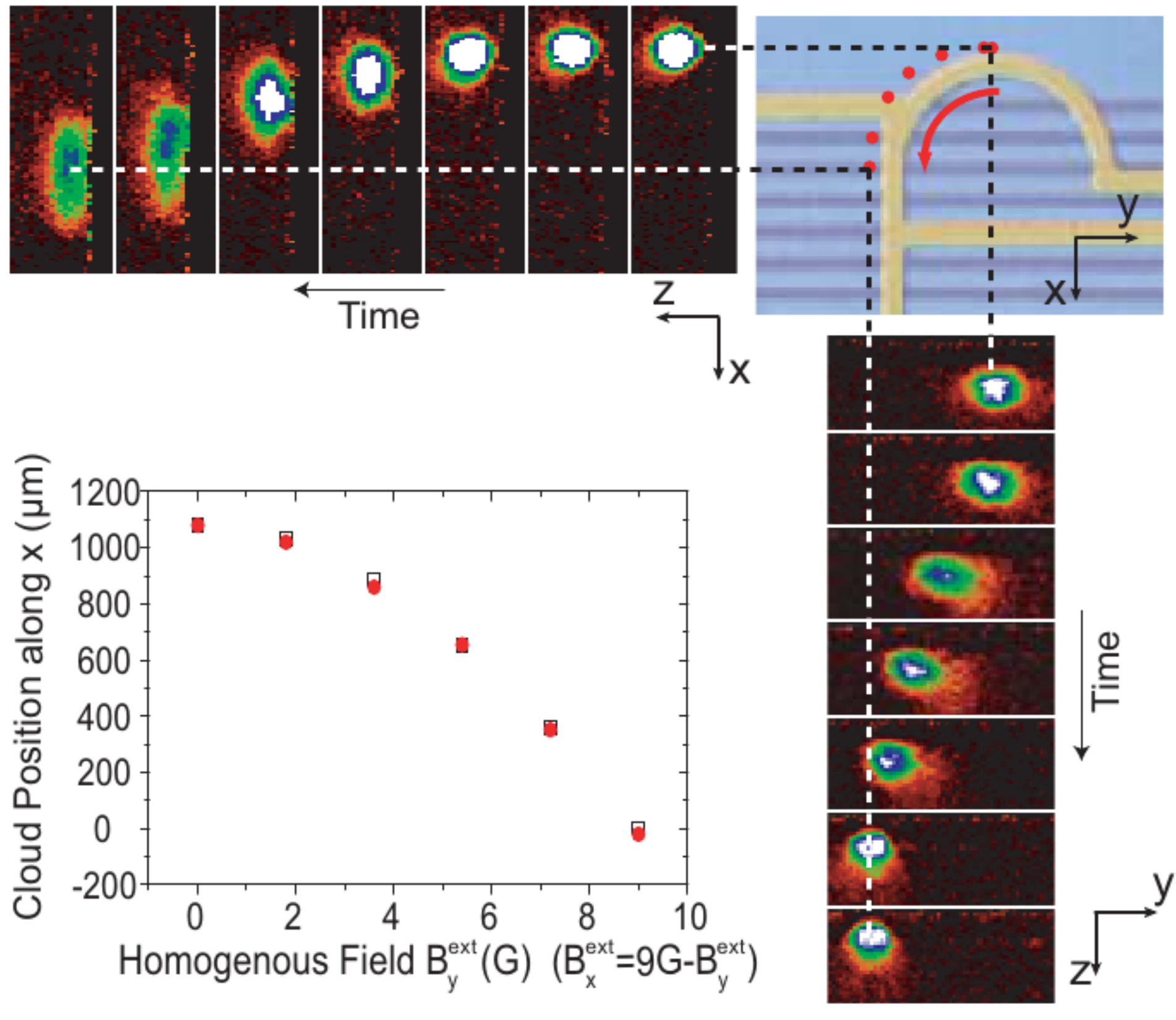}
 \caption{Illustration of positional precision  that can be realized by using an atom chip. The figure shows the 
 rotation of a cloud of trapped \(^{87}\)Rb atoms. The absorption images were acquired along two perpendicular axes 
 parallel to the substrate. The pictures are scaled so that  the wire dimensions may serve as a length scale for  
 determining the cloud size. Reprinted with permission from Long {\it et al.,}~\cite{long} \copyright\ 2005 by Springer Eur. Phys. J. D.}
 \label{fig:guideconv}
 \end{center}
\end{figure}

Atom chips offer a convenient and flexible way for precise spatial positioning of the atom clouds at the input port of the 
interferometer.  Long {\it et al.} used a magnetic conveyor  to transport  atoms over a total distance 
of 24~cm.~\cite{long}  Figure~\ref{fig:guideconv} shows the operation of this conveyor, which rotates the atom cloud 
position by 90\(^\circ\).
Another relevant development for guided atom interferometry is their demonstration of a magnetic guide that does not require  an external 
bias field. The guide is loaded by the rotating-trap device shown in Fig.~\ref{fig:guideconv}. In other work, G\"unther 
{\it et al.} demonstrated  sub-micron accuracy in detecting the position of a Bose--Einstein condensate,~\cite{gunther} which  enabled a 
demonstration of magnetic-field microscope sensing at the mG level.

\begin{figure}[htb]
 \begin{center}
 \includegraphics[width=0.8\columnwidth]{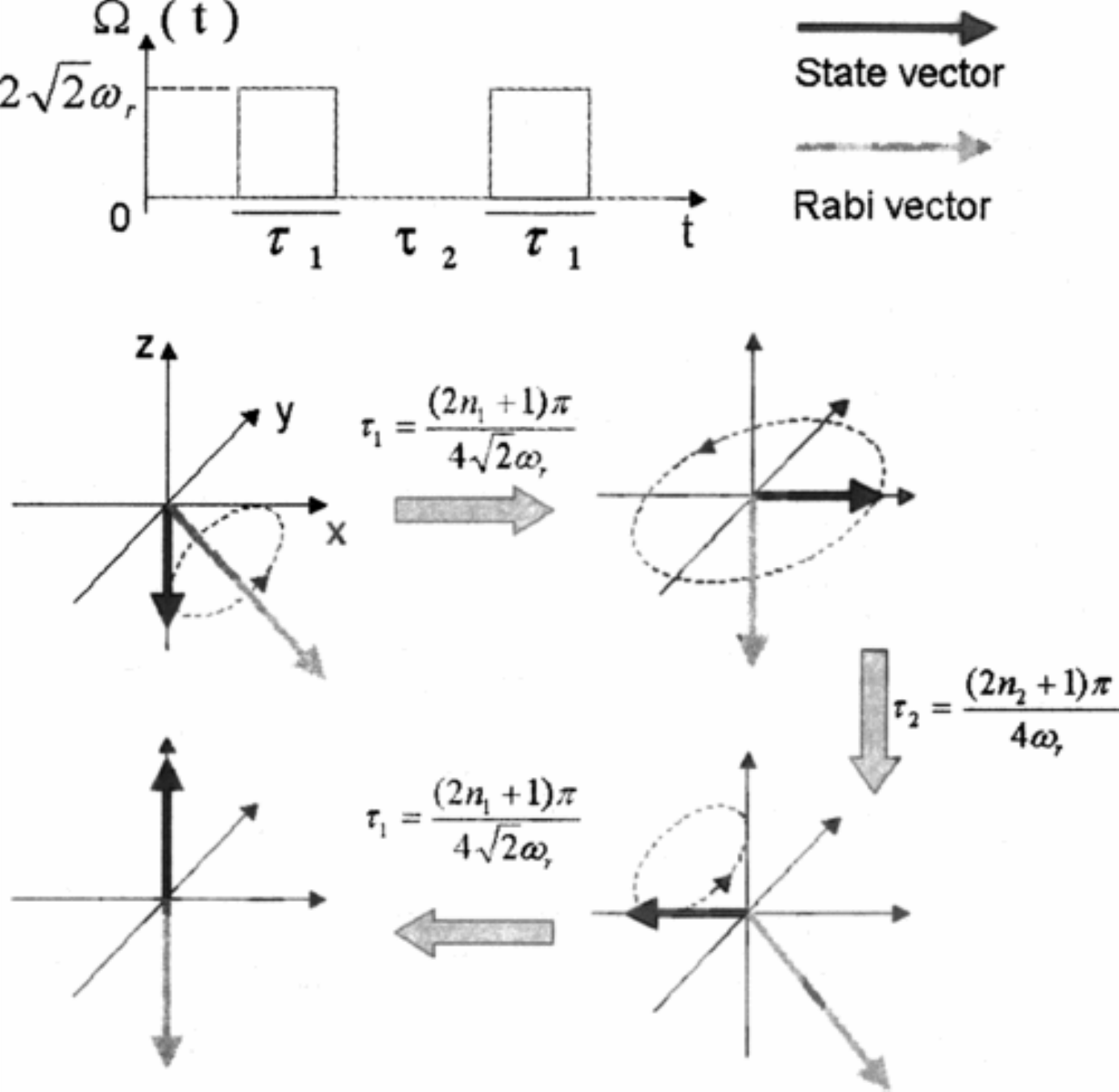}
 \caption{(top) Illustration of the two square-shaped pulse sequence. (bottom) Representation of the momentum-state vector in 
 the Bloch sphere. Adapted from Wu {\it et al.},~\cite{swu} \copyright\ 2005 by the American Physical Society.}
 \label{fig:pm2hbk}
 \end{center}
\end{figure}

\begin{figure}[htb]
 \begin{center}
 \includegraphics[width=\columnwidth]{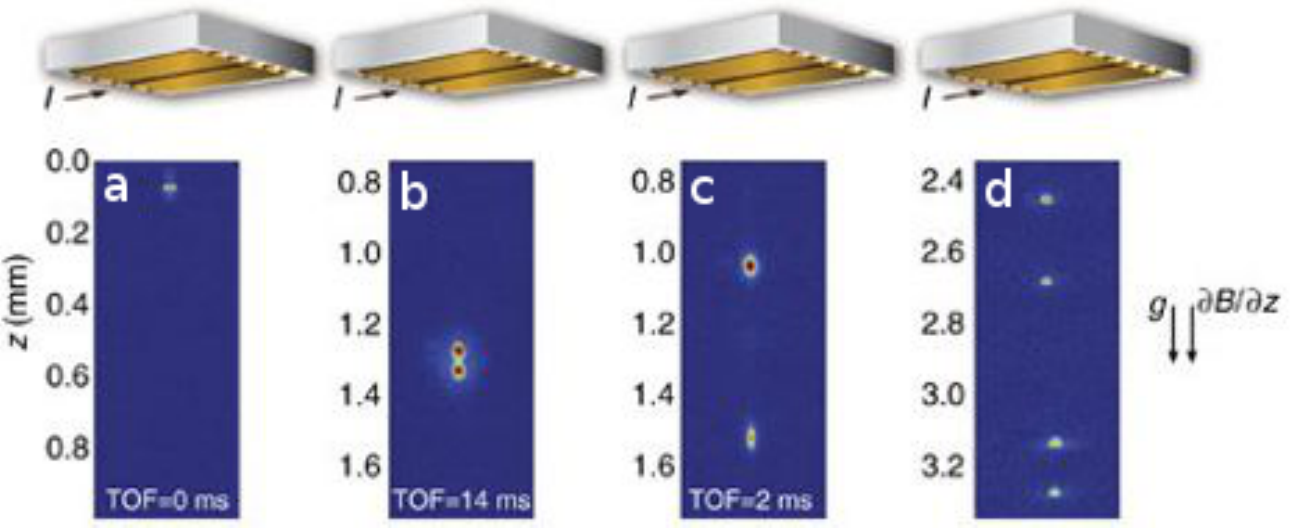}
 \caption{Time-of-flight absorption images  (a) of the trapped cloud. In panel (b), the momentum splitting 
 is weak (less than $\hbar k$) when using 5~\(\mu\)s of interaction time with the magnetic-field gradient 
 (\(\text{TOF}=14\)~ms). The  splitting realized in panel (c) corresponds to $40\hbar k$ of momentum transfer using 1~ms interaction time 
 and 2~ms  TOF. In panel (d), the absorption images show when the internal states are separated by an additional strong 
 gradient pulse. Adapted from Machluf {\it et al.},~\cite{machluf} \copyright\ 2013 by Springer Nature Communications.}
 \label{fig:sgerl}
 \end{center}
\end{figure}

Different atom interferometer techniques are available to launch  atoms into the guide by 
subjecting them to a momentum kick. For 
magnetically guided atoms, the standard solution is to use  double Bragg diffraction.~\cite{swu,giese} In particular, 
Wu {\it et al.} developed a theory to describe a pulse sequence of two square-shaped standing-wave light 
pulses.~\cite{swu} By properly choosing the strengths, durations, and separation between pulses, a Bose--Einstein condensate at rest may be almost 
perfectly (99\%) split into the \(\pm 2\hbar k\) diffraction orders.~\cite{wang,swu} Figure \ref{fig:pm2hbk} shows the designed 
pulse sequence.
On atom chips, coherent momentum splitting can be produced by using a Stern--Gerlach beam splitter, as demonstrated by Machluf 
{\it et al.} in Ref.~\onlinecite{machluf}. The beam splitter results from a combination of magnetic-field gradient pulses
and RF transitions between Zeeman states. Both the gradient and the RF field are generated by feeding current to on-chip 
wires.  Figure~\ref{fig:sgerl} shows absorption images of the splitting process.
In this work the authors demonstrate interference fringes characterized by short-term phase fluctuations of 
about 1 radian and 
long-term drifts on the time scale of an hour. State-dependent microwave potentials can also be used to induce momentum 
transfer with on-chip microfabricated wires.~\cite{bohi}

Usually, the number of atoms after  application of the final beam-splitter pulse is used to determine the inertial phase 
at the end of a single AI cycle. This is done for all the propagation modes in order to compute the associated 
probabilities, as discussed in Sec.~\ref{sec:perf}. The standard technique is based on absorption imaging, as already 
presented, for instance, in Fig.~\ref{fig:guideconv}. However,  locally detecting  atoms 
at  well-defined positions on the chip is advantageous 
for three main reasons: The first reason is that individual manipulation of the atom clouds (or single atoms) becomes possible, which can 
be useful for preparing one atom cloud and, {\it at the same time}, 
measure the AI output on the previously prepared cloud. This would allow, for example, an 
interleaved 
operation configuration.~\cite{savoie} The second reason is the possibility to implement  local quantum 
non-demolition measurements of  atom number at the AI output. Such a strategy would allow recycling of  atoms in the 
interferometer to increase its measurement bandwidth. Finally, the third reason is 
that integrated atom detection can be implemented (for instance, 
{\it in situ} collection of  fluorescence  by using an on-chip fiber), which is important for making compact sensors and 
avoiding the use of  CCD cameras.

\begin{figure}[htb]
 \begin{center}
 \includegraphics[width=0.8\columnwidth]{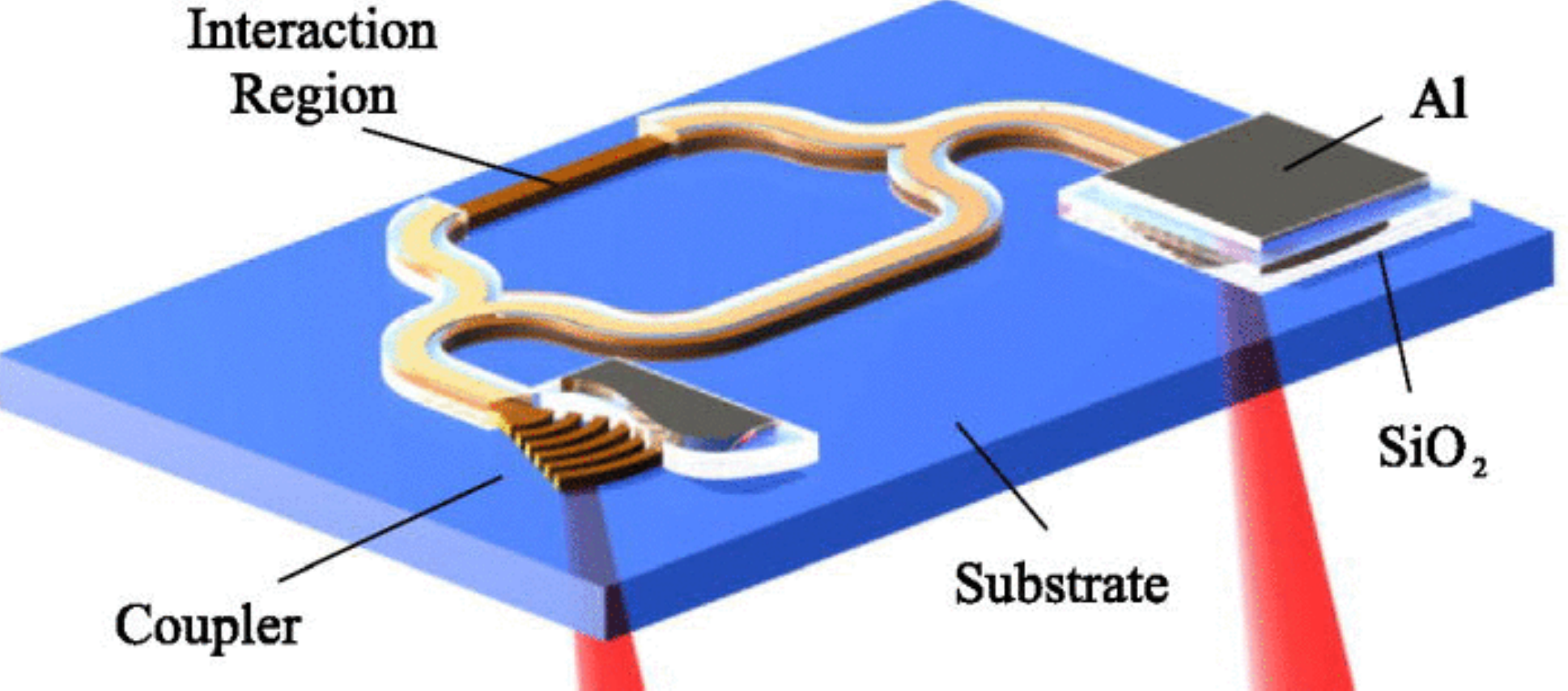}
 \caption{Schematic drawing of an integrated Mach--Zehnder optical interferometer containing an interaction or detection region for 
 the atoms in one of its arms. Light is coupled in by focusing a laser beam onto a grating coupler. Another grating 
 coupler is used to extract the light from the device. Reprinted with permission from Ritter {\it et al.},~\cite{ritter} 
 \copyright\ 2015 AIP Publishing LLC.}
 \label{fig:coupleurint}
 \end{center}
\end{figure}

\begin{figure}[htb]
 \begin{center}
 \includegraphics[width=0.8\columnwidth]{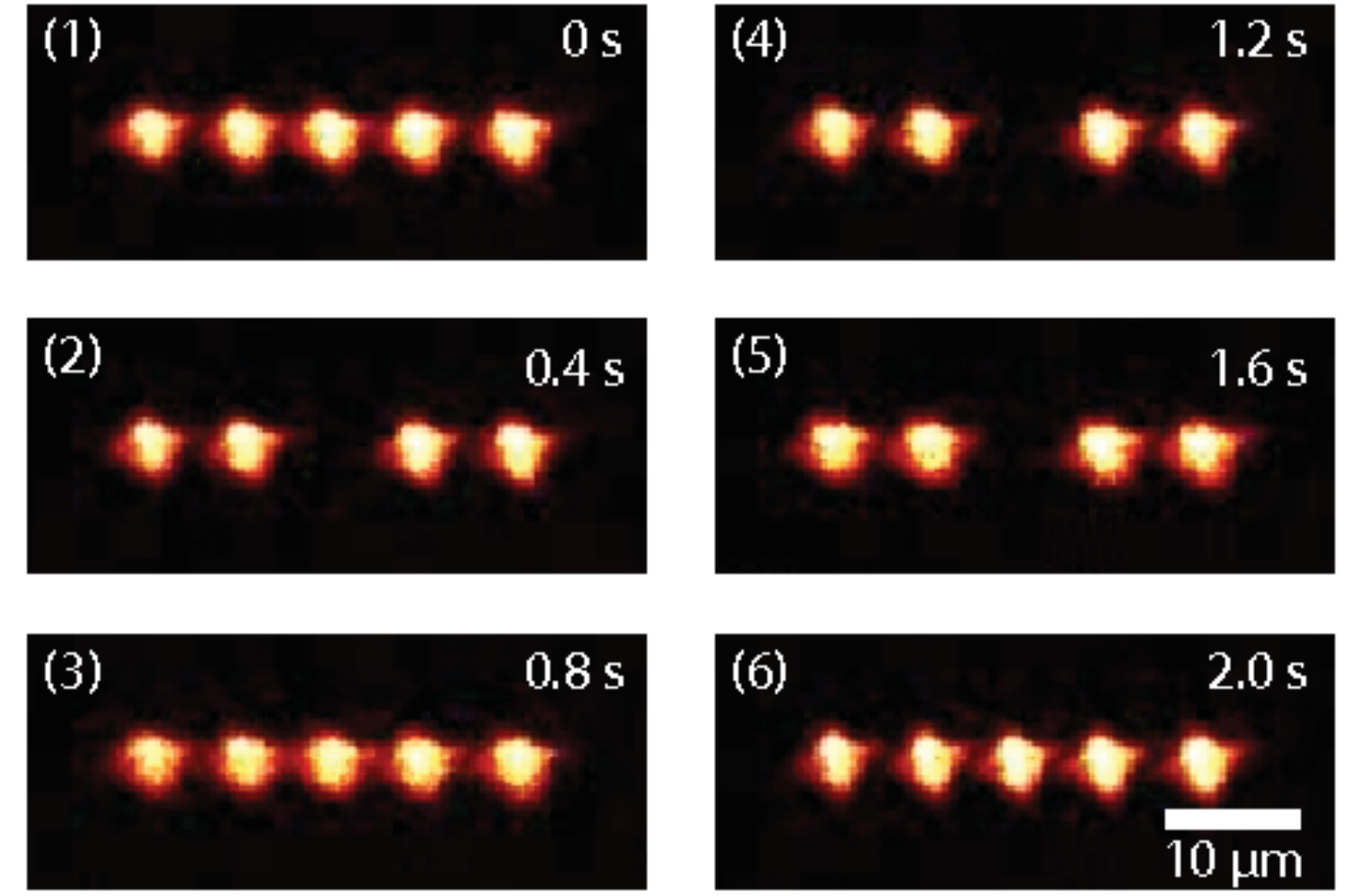}
 \caption{Fluorescence images showing single-ion addressing. The grating coupler is aligned with the central ion, which 
is  prepared in a dark state by focusing a 674 nm laser  onto the grating. The fluorescence  is at 422 nm, and  
the signal is  collected for 2~s. Adapted from Mehta {\it et al.},~\cite{mehta1} \copyright\ 2016 by Springer 
Nature Publishing AG.}
 \label{fig:sqbitaddr}
 \end{center}
\end{figure}

The enabling technology in this case has already been demonstrated by using integrated optical elements on an atom 
chip.~\cite{wilzbach} Another solution is to combine an atom chip with integrated photonic optical 
chips. In both situations, atoms may be locally excited for position-resolved 
detection.~\cite{ritter,mehta,mehta1,kohnen}  Figure~\ref{fig:coupleurint} shows a schematic representation of an optical chip used to 
acquire the spectrum of a hot vapor of rubidium atoms. It is based on a Mach--Zehnder light interferometer with one  
 arm exposed to the atomic vapor. The evanescent field of the optical waveguide mode interacts with the 
atoms, and the
interaction strength is read out as a modification of the interferometer transmission. In other words, the presence of the 
atoms produces a phase shift that carries information on their number and internal state.
In another experiment, Mehta {\it et al.} used a grating coupler to implement position-resolved excitation of 
ions~\cite{mehta1} and  addressed  a single \(^{88}\)Sr\(^+\)  qubit, 
as shown in Fig.~\ref{fig:sqbitaddr}.

\begin{figure}[h!]
 \begin{center}
 \includegraphics[width=\columnwidth]{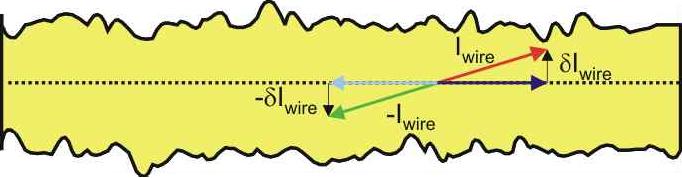}
 \caption{Top view of  current \(I_{\rm wire}\) flowing in a wire with corrugated walls.}
 \label{fig:rugo}
 \end{center}
\end{figure}

To guarantee monomode propagation of guided atoms~\cite{golding,andersson,burke1,das,wu1} when realizing  magnetic 
waveguides with relatively low currents 
(below hundreds of mA),  the atoms must be confined close to the chip surface. However, in this situation, the coherent 
properties of the atomic states can be dramatically affected by the corrugation of the microwires. These fabrication 
defects produce a magnetically rough potential that can destroy the system coherence or, even worse, blockade 
the propagation of the atoms.~\cite{kruger,Whitlock,sinclair:031603} Fabrication defects have been a fundamental limitation for atom 
chips.~\cite{kraft,Hinds,leanhardt:100404,EstevePRArug,lukin,sculturing,EsteveDWell} The physical origins of this 
roughness can be understood as follows: Consider a wire with a current flowing  left to right as shown 
by the dark blue arrow in Fig.~\ref{fig:rugo}.
Because of meanders  in the wire's wall,  an electron flux exists in the direction perpendicular to the wire. These 
electrons  generate a current \(\delta I_{\rm wire}\) that in turn produces a magnetic trapping potential in the direction 
parallel to the wire.

\begin{figure}[htb]
 \begin{center}
 \includegraphics[width=\columnwidth]{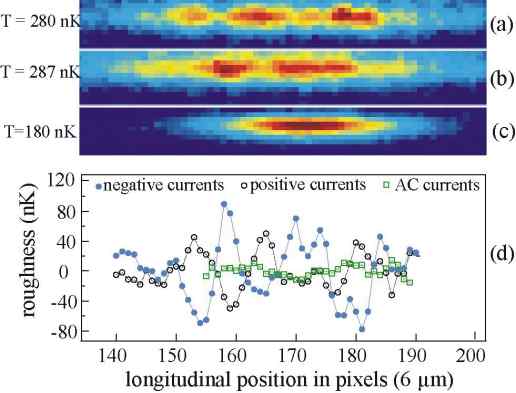}
 \caption{(a) Absorption images of a thermal cloud for negative DC, (b) positive DC, and (c)\ AC currents. Plotted is the number of atoms per pixel (pixel size is \(6 \times 6~\mu\)m\(^2\)). Panel (d) shows the potential 
 roughness  extracted from longitudinal profiles by using the Maxwell--Boltzmann distribution. Adapted from 
 Trebbia {\it et al.},~\cite{trebbia} \copyright\ 2007 by the American Physical Society.}
 \label{fig:rugo1}
 \end{center}
\end{figure}

Figure \ref{fig:rugo1}(a) shows an in-trap absorption image of a cold thermal \(^{87}\)Rb cloud that exhibits fragmentation. This 
image corresponds to atoms loaded in a linear magnetic guide created with three microfabricated wires,~\cite{trebbia} each of which 
carried a DC current in a configuration that creates a quadrupole trap close to the minimum of the guiding potential. When 
the current directions are reversed, the extrema of the rough magnetic potential are also reversed, as can be seen in 
Fig.~\ref{fig:rugo1}(b). However, the atoms are now confined in complementary positions with respect to Fig.~\ref{fig:rugo1}(a), and
it is this complementarity that inspired the modulation technique that  suppresses the 
roughness of the magnetic 
guide.~\cite{trebbia} In fact, by temporally modulating the current in the 
wires, Trebbia {\it et al.} demonstrated a 
roughness-free guide, as can be seen in the absorption image shown in Fig.~\ref{fig:rugo1}(c). The rough potential was 
extracted by fitting the linear density 
with a Maxwell--Boltzmann distribution along the axis of the cloud in Figs.~\ref{fig:rugo1}(a)--\ref{fig:rugo1}(c).~\cite{bouch} This result is presented in Fig.~\ref{fig:rugo1}(d), which  quantifies the fragmentation complementarity 
observed in the previous figures and the degree of suppression of the roughness. For this technique to work, two 
criteria must be satisfied: (a) the modulation frequency must exceed the radial guide frequency but not 
the Zeeman frequency; (b) the phase difference between the currents must be constant.
 
\begin{figure}[htb]
 \begin{center}
 \includegraphics[width=0.8\columnwidth]{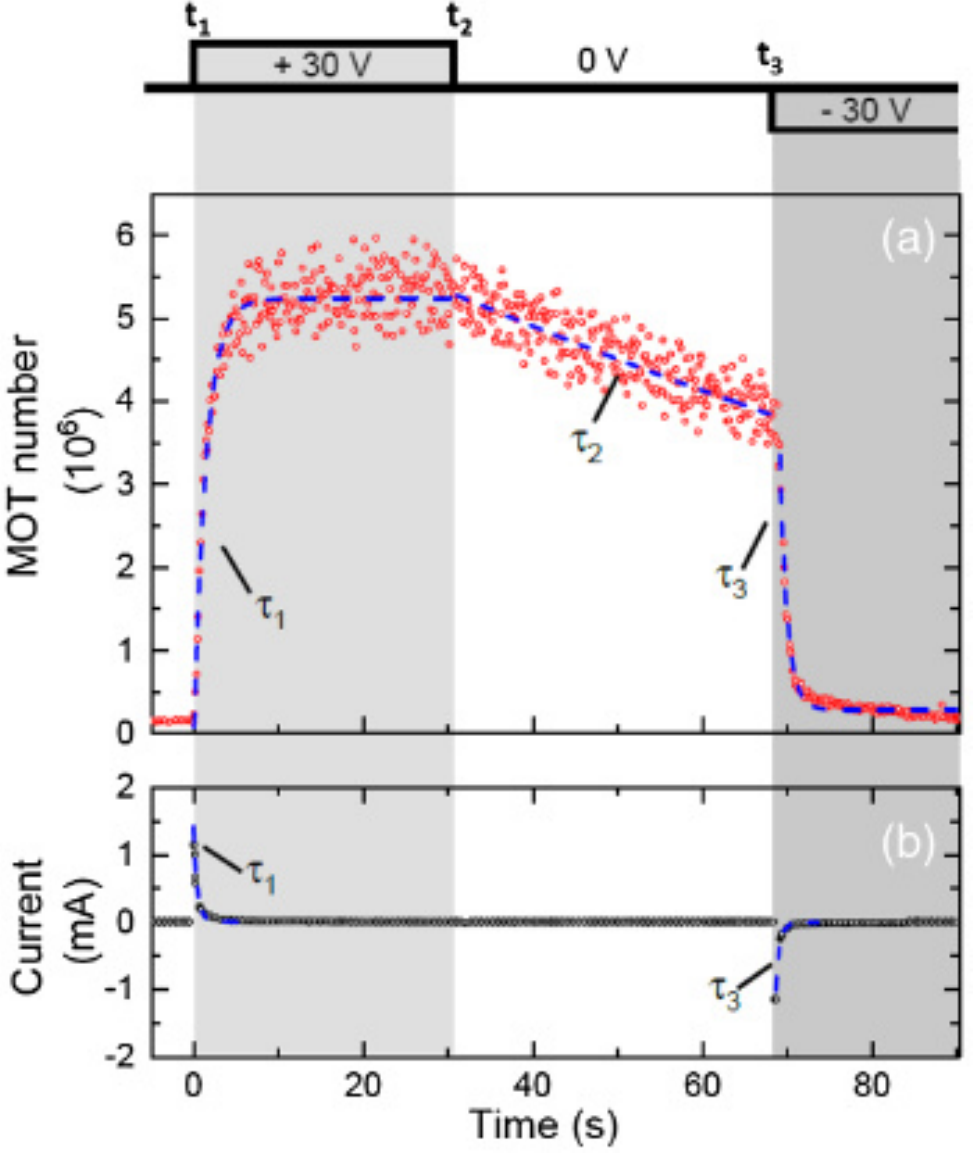}
 \caption{(a) MOT time response is represented by  experimental points (red dots), which correspond to the measured 
atom number. (b) The current flowing in this reversible source is given by the black points. Adapted from Kang 
{\it et al.},~\cite{kitching} \copyright\ 2019 by the Optical Society of America.}
 \label{fig:aliminvac}
 \end{center}
\end{figure}

One of the key components of any cold atom sensor is the vacuum system. The simplest vacuum setup would use a single chamber 
incorporating the atom source (i.e.,\ an alkali-metal dispenser). However, 
this solution requires a highly dynamic vacuum system capable of switching from high (\(
{\approx}10^{-8}\)~mbar) to low 
pressure (\(
{\approx}10^{-11}\)~mbar) in a few tenths of milliseconds.~\cite{Dugrain} Such a response is needed  to reduce the 
deleterious effect of the dead time between measurements. Recently, a promising technique was demonstrated~\cite{kitching} 
that allows one to quickly and reversibly control the Rb background pressure in a cell, with switching times close to 1~s. This reversible operation is illustrated in Fig.~\ref{fig:aliminvac}, where the magneto-optical-trap (MOT) 
loading and depletion are controlled 
by a voltage applied across the electrodes.

Other relevant investigations on the optimized operation of compact ultrahigh vacuum (UHV) systems have been 
reported in Ref.~\onlinecite{campo}. In this work, light-induced atomic desorption is used to modulate the background pressure 
of \(^{87}\)Rb atoms in a glass cell. For atom interferometry and atom sensors, a UHV system was designed and tested 
for operation in the highly vibrating environment of a rocket.~\cite{grosse} In Ref.~\cite{scherer} the authors 
investigated the use of passive vacuum pumps (non-evaporable getter pumps) for the development of compact cold 
atom sensors. Basu and Vel\'asquez-Garcia~\cite{basu} demonstrated microfabricated non-magnetic ion pumps   to 
maintain UHV conditions in miniature vacuum chambers for atom interferometry. Finally, Martin \textit{et al.}~\cite{jmm} investigated the pumping 
dynamics of sputtered ion pumps. This study  describes the main physical pumping 
mechanisms 
based on a nonlinear model for the ion-current pump-down dynamics in the low-pressure regime. The results  suggest
that a three-dimensional (3D) structured cathode might allow for a smaller pump, increasing or at 
least holding constant the trapping 
cross section of this electrode, which may have significant consequences on the design and development of miniature 
ion pumps.

\section{Case study of a sensor design}\label{sec:perf}
This section presents a case study of a rotation sensor design that uses an atom chip. The working principle is based on the Sagnac effect, which naturally suggests a circular guiding geometry for the AI.

\subsection{Measuring rotation with a waveguide}
\begin{figure}[htb]
 \begin{center}
 \includegraphics[width=0.9\columnwidth]{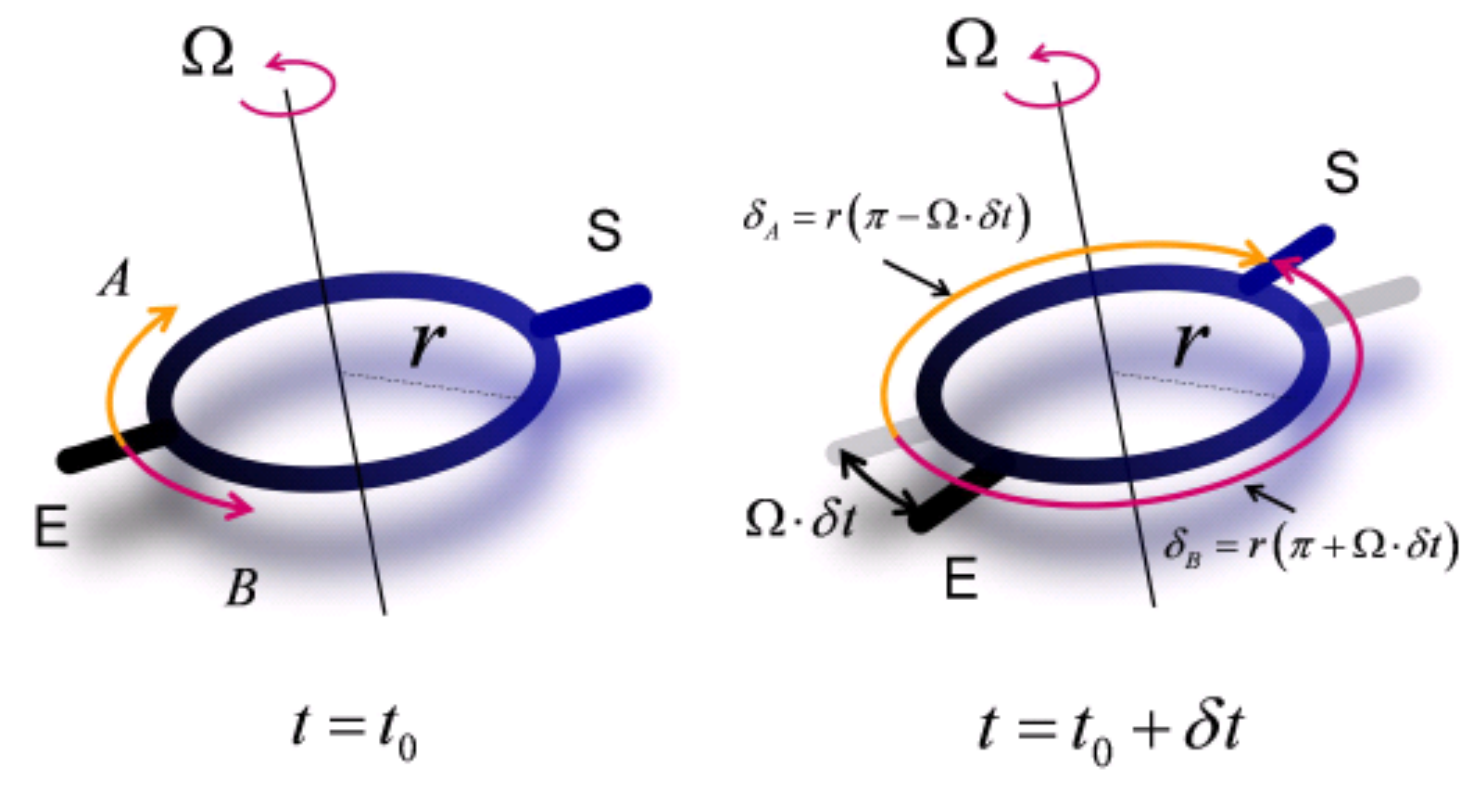}
 \caption{Illustration of the Sagnac effect in a circular guide.}
 \label{fig:gyrcirc}
 \end{center}
\end{figure}

To establish the physics  that will drive the gyrometer design,  consider the Sagnac effect as sketched in 
Fig.~\ref{fig:gyrcirc}. Starting with the illustration on the left side of the figure,  assume that, at \(t=t_0\), 
two particles \(A\) and \(B\) leave the AI entry port \(E\) and  propagate freely in the azimuthal direction of 
this circular guide of radius \(r\). If the guide is rotated about an axis \(
{\bf \Omega}\) perpendicular 
to the guide's oriented area, then after a time interval \(\delta t\)  particles \(A\) and \(B\) would  travel  
a geometrical path length \(\delta_A = r(\pi-\Omega \delta t)\) and \(\delta_B = r(\pi+\Omega \delta t)\), respectively. Therefore, there  exists 
a path-length difference \(\delta_{AB}=\delta_B - \delta_A\) determined by the different arrival times of the particles 
at the exit port \(S\). By measuring this quantity, we can determine the magnitude and direction of the rotation of the apparatus. 

In the following, we discuss the key elements  of the design of an atom-chip-based cold atom gyrometer. To be 
quantitative, we consider magnetic guiding of ultracold thermal \(^{87}\)Rb atoms.

\subsection{Sensor design}
\subsubsection*{Guiding}
\begin{figure}[htb]
 \begin{center}
 \includegraphics[width=6.25cm]{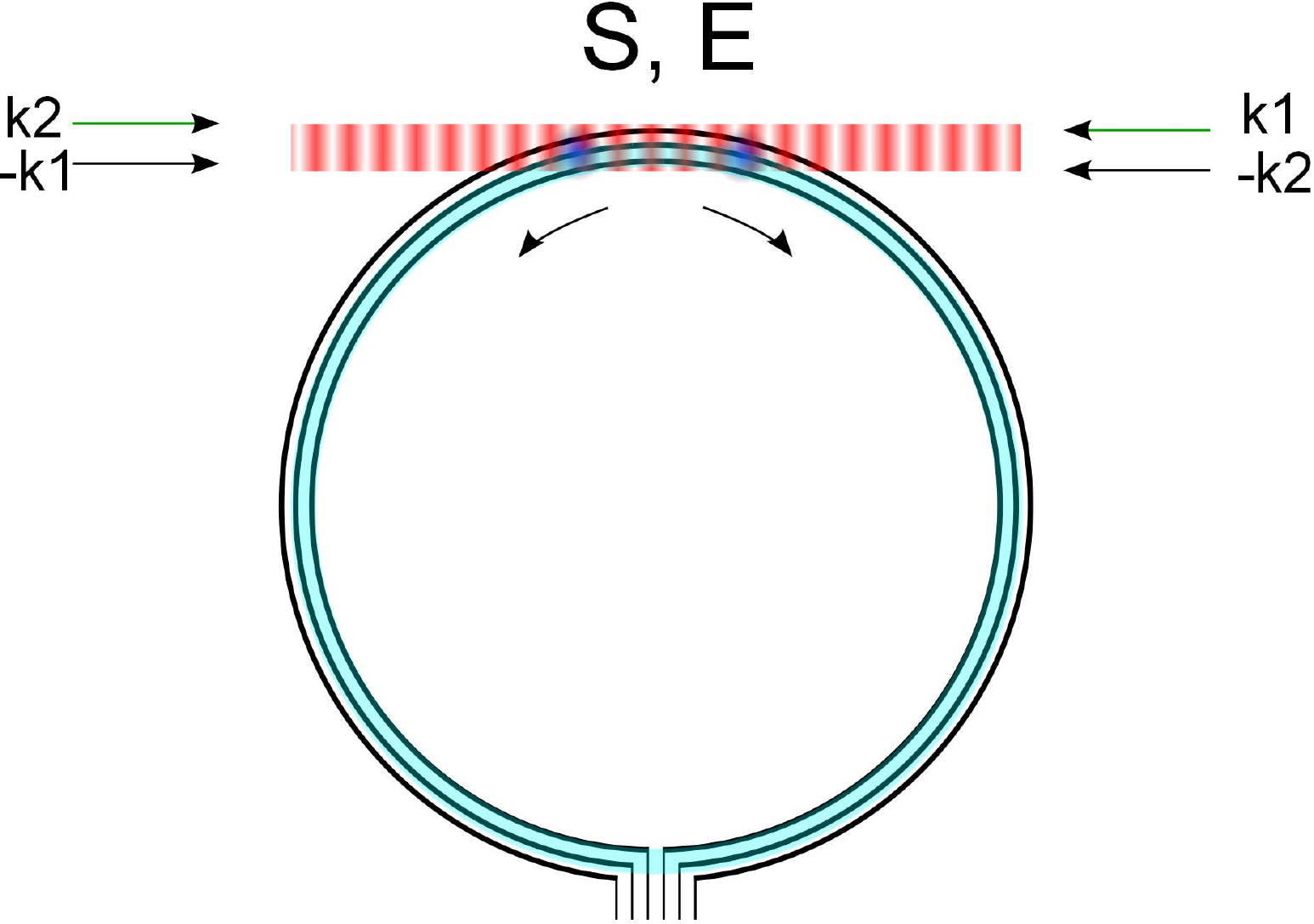}
 \caption{Microfabricated-wire configuration used to generate the circular magnetic guide shown in Fig.~\ref{fig:guide}. Here, 
 the entry \(E\) and output \(S\) ports are located at the same point in space. The laser beams implementing the matter-wave 
 optics are represented by the wave vectors \(
{{\bf k}_1}\) and \(
{{\bf k}_2}\).} 
 \label{fig:3w}
 \end{center}
\end{figure}

\begin{figure}[htb]
 \begin{center}
 \includegraphics[width=7.5cm]{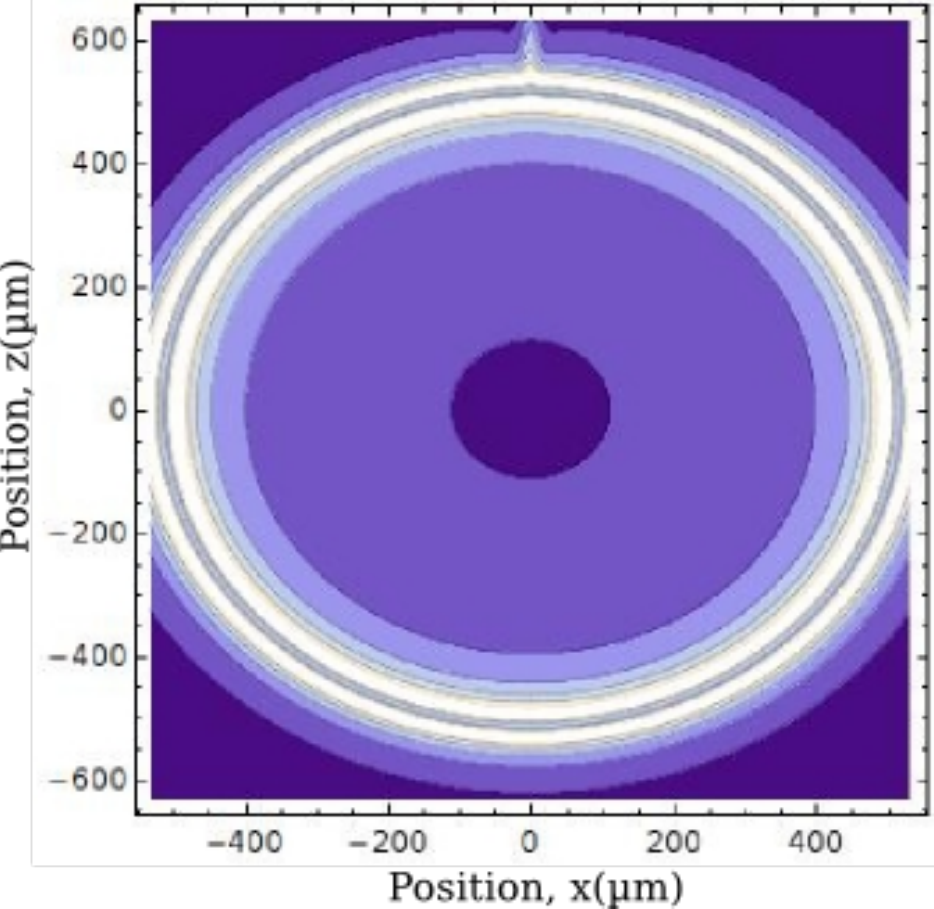}
 \caption{Magnetic guiding potential for  on-chip cold atom gyrometer. It is produced by a three-wire configuration in 
which the distance between the wires is  13~\(\mu\)m and the central circular wire has a radius of 500~\(\mu\)m.}
 \label{fig:guide}
 \end{center}
\end{figure}

As we have  seen,  several solutions exist for realizing a matter-wave guide. To implement an inertial sensor and exploit the potential 
for compactness offered by an atom chip,  we  consider a circular 
magnetic guide produced by three on-chip microfabricated concentric wires. Figure  \ref{fig:3w} shows the wire pattern, which does not require an external bias field produced with coils. This is a well-known configuration in which 
the external wires generate the bias magnetic field needed to cancel the magnetic field of the central wire.
The resulting field is a quadrupole field and, by modulating the current fed to the wires, the atoms can be guided at the minimum 
of this field in an equivalent roughness-free stable magnetic guide.~\cite{trebbia,clgapra} Note that the entry \(E\) and 
output \(S\) ports are located at the same point in space. Such 
a geometry allows rejection of common-mode noise.  Figure~\ref{fig:guide} shows the magnetic potential generated by this configuration 
of microfabricated wires. This magnetic potential is obtained when the wires are separated by 13~\(\mu\)m and carry modulated currents 
of \(-123\) mA and 121 mA in the external and central wires, respectively. The radius of the central wire is 
  \(R=500~\mu\)m. The magnetic guide obtained in this way is around 13~\(\mu\)m from the chip surface.
Given these parameter values, the radial guiding frequency is about 1.5 kHz, and the potential 
depth is roughly 300~\(\mu\)K. These guide design parameters allow the propagation of cold thermal 
atoms. By using this type of source we can avoid the important contribution of atom-atom interactions to the 
systematic noise and drift in the inertial phase of interest,~\cite{wang,horikoshi,jo,li,stickney} and 
less preparation time is required than with 
Bose--Einstein-condensed samples. In fact, concerning just the cooling process, cold thermal atoms used in atom interferometry 
are obtained after molasses cooling whereas degenerate samples require an additional  evaporative-cooling 
step. However, quantum 
degenerate 
atom sources might be relevant when realizing, for instance, large-momentum-transfer beam 
splitters.~\cite{mcdonald,debs} Actually, as  discussed later, the sensitivity of the interferometer can be 
enhanced 
by increasing the launch velocity and by letting the atoms undertake several trips around the guide at a fixed single-loop 
interrogation time \(2T\).

\subsubsection*{Guide fabrication}
The realization of the proposed configuration, shown in Fig.~\ref{fig:3w}, raises 
several fabrication challenges. For instance, the fabrication of the microwires must account for the relevant practical 
and physics 
constraints of a compact sensor for inertial navigation. From the practical point 
of view,  power consumption is another  relevant constraint (for instance, to operate the sensor on batteries). As already 
discussed, this problem can be mitigated by guiding the atoms close to the chip surface, at a few tens of 
microns. In this case, the proximity to the chip allows the  production of strong magnetic-field gradients to confine 
the atoms by using currents well below 1 A.~\cite{note2} This point is important for two main reasons: First,  
reducing the current intensity reduces the heat dissipated by the wires. Second,  developing or adding a heat-management 
solution to the sensor, which can be 
detrimental to its performance, is not necessary (e.g., a cooling system that introduces parasitic vibrations). However, from the physics point of view, as already mentioned, the proximity to the 
wires renders the guided atoms sensitive to the roughness of the magnetic potential. This problem suggests a microfabrication 
process based on metal evaporation, but this is a costly solution when considering mass production of these sensors. The 
combined challenge of low roughness and low cost can be addressed by using a metallization process based 
on electroplating. The only requirement that remains in this case is the realization of relatively small metal grains.~\cite{lev} In addition, using an electroplating process allows  the fabrication of microwires with cross sections with 
large aspect ratios. In this way, their electrical resistance and, consequently,  heat dissipation can be reduced.

\begin{figure}[htb]
 \begin{center}
 \includegraphics[width=8.12cm]{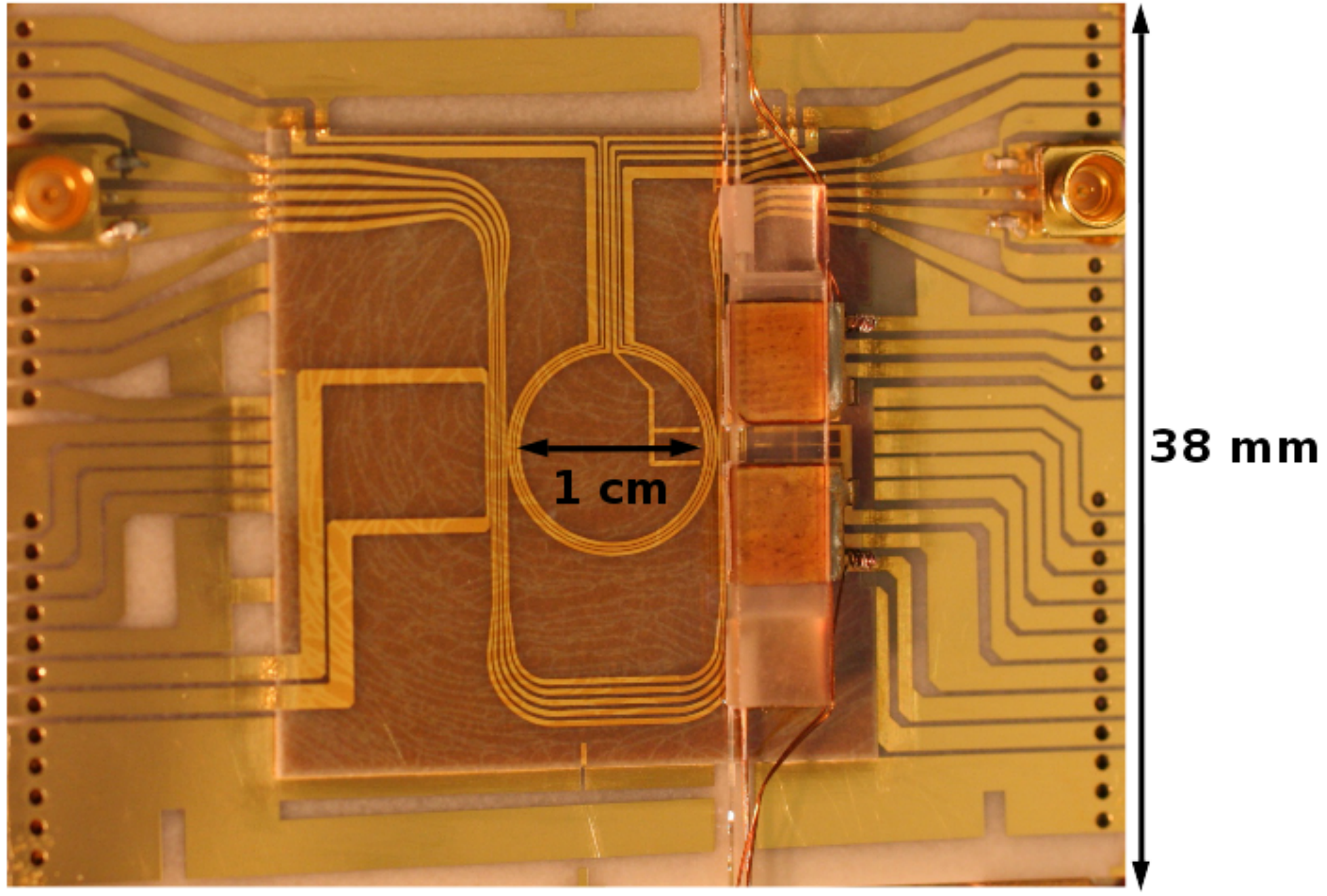}
 \caption{Photograph of the atom chip of the LKB-SYRTE experiment Trapped Atom Clock on a Chip. The experiment is 
 developing an entanglement-enhanced atomic clock with trapped atoms on a chip. More details are available in Ref.~\onlinecite{huang}. Courtesy of J. Reichel (Paris).}
 \label{fig:eetacc}
 \end{center}
\end{figure}

Figure~\ref{fig:eetacc} shows an example of a circular magnetic guide used to transport atoms 
from the laser-cooling region to the science region of an atom chip.~\cite{huang} The picture illustrates a three-wire circular 
geometry of 1~cm diameter in 
which the atoms are transported at 240~\(\mu\)m from the atom chip surface, which demonstrates the feasibility of fabricating the wire 
pattern needed to produce circular magnetic guiding potentials on an atom chip. The bottom chip is used to realize an 
evaporative cooling trap and also serves as an electrical feedthrough for the in-vacuum wires of the top chip in which the 
circular trap is implemented. Both atom chips were fabricated by using electroplating, which is a well-established technology in the 
microelectronics industry.

\subsubsection*{Scale factor}
\begin{figure}[htb]
 \begin{center}
 \includegraphics[width=8.12cm]{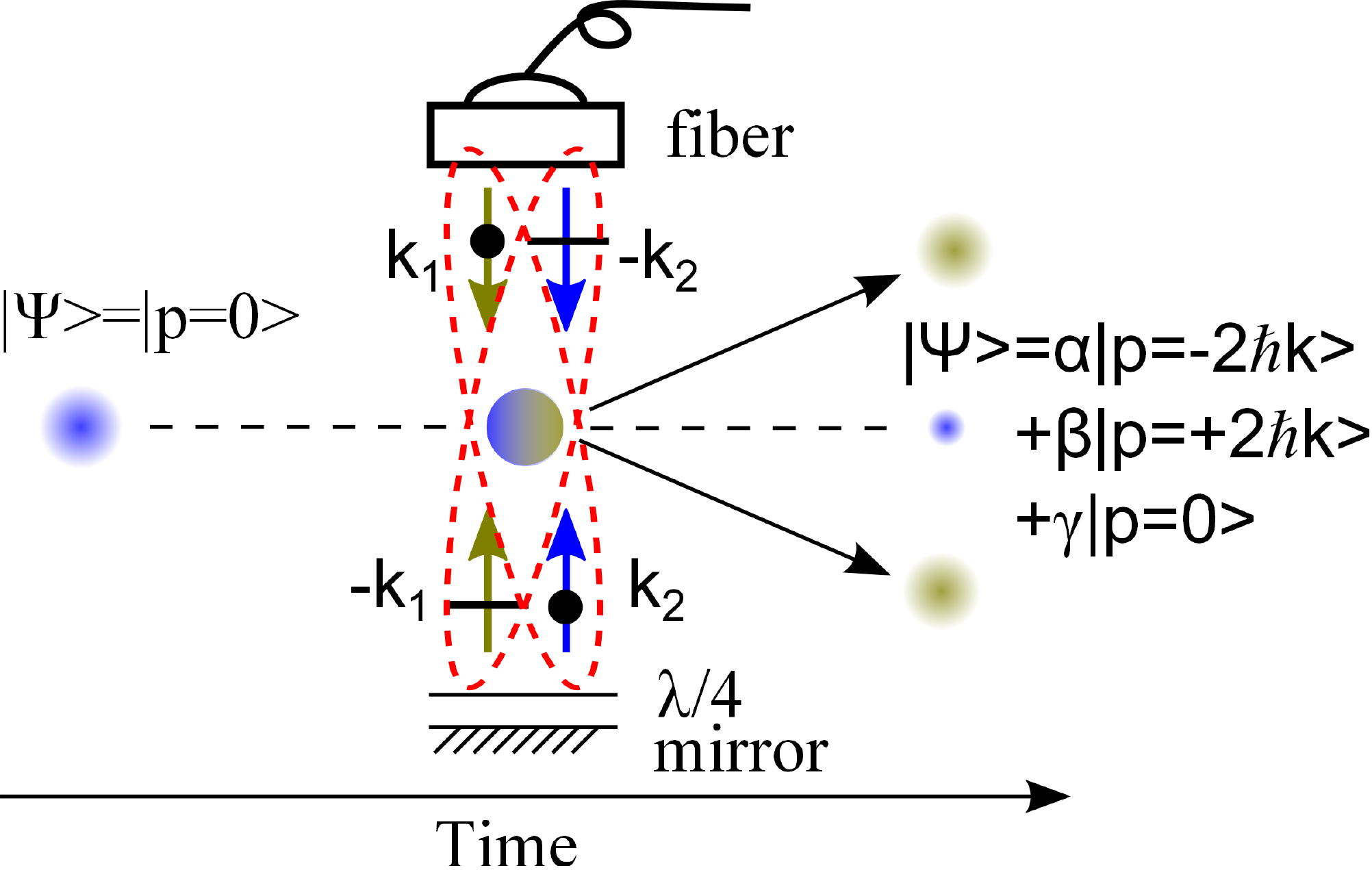}
 \caption{Beam-splitting process for the circular guiding potential considered in Fig.~\ref{fig:3w}. It represents a 
double Bragg diffraction,~\cite{ahlers} each  realized by the pair of beams enclosed by a given red dashed ellipse.}
 \label{fig:braggsep}
 \end{center}
\end{figure}

When using magnetically guided atoms, it is preferable to use Bragg 
transitions~\cite{giltner,ahlers} to realize 
the beam-splitter and mirror light pulses.~\cite{moler} In fact, in a Bragg transition, only the atomic momentum 
is modified and the atoms remain in the same magnetic-trappable internal state. In a circular waveguide, only beam-splitter 
pulses need  be implemented because 
the atomic trajectories are, by construction, deflected by the guide. For the geometry shown in Fig.~\ref{fig:3w}, we  
require only two Bragg transitions, as indicated in Fig.~\ref{fig:braggsep}. If we do not use a composite pulse 
sequence,~\cite{swu} the initial state \(|\Psi\rangle=|{\bf p}=0\rangle\) is transformed into a three-component superposition state
\(|\Psi\rangle=\alpha|{\bf p}=-2\hbar {\bf k}\rangle + \beta|{\bf p}=+2\hbar {\bf k}\rangle + \gamma|{\bf p}=0\rangle\). 
By using the path-integral approach,~\cite{pippa} one can show that the relevant inertial (rotation) phase defining the 
scale factor is given by the expression
\begin{equation}
 \Phi = 4 k R \Omega (2T) = 2\times2\frac{M}{\hbar} A \Omega,
\label{eq:echfac}
\end{equation}
where \(2T=\pi R/v_{\rm r}=M\pi R/(\hbar k)\) is the interferometer interrogation time for a round trip in a guide of area
\(A=\pi R^2\) and \(v_{\rm r}\) is the recoil velocity. In Eq.~(\ref{eq:echfac}), we have considered an ideal beam-splitting 
process to compute the probability of finding the atoms at the output port \(|{\bf p}=0\rangle\) of the AI.

\subsubsection*{Rotation-rate sensitivity} 
When the AI is operated at mid-fringe (maximum phase sensitivity), the probability \(P(\Omega)\) of finding the atoms at the output port 
\(|{\bf p}=0\rangle\) is  
\begin{equation}
 P(\Omega)=\frac{N}{2}\big[1-\eta\cos\big(\Phi+\pi/2\big)\big],
\label{eq:pp0}
\end{equation}
 where \(N\) and \(\eta\) are  the number of atoms and the contrast of the AI, respectively.  Equation~(\ref{eq:pp0}) 
states that an infinitesimal variation of \(P(\Omega)\) translates into an infinitesimal variation of the rotation rate such that
\begin{equation}
 \delta \Omega=\frac{1}{\Big|\frac{d P(\Omega)}{d\Omega}\big|_{\Omega=0}\Big|}\delta P(\Omega=0).
\label{eq:sensiom}
\end{equation}

In Eq.~(\ref{eq:sensiom}), \(\delta P(\Omega=0)\) gives the probability noise around the working point of the AI,  
which can be chosen, for example, by setting a fixed phase difference between the Bragg 
beams. For instance, in Eq.~(\ref{eq:pp0}) the phase difference is  \(\pi/2\), such that \(P(\Omega)\) is  the half-fringe value 
of the interference signal in the 
absence of rotations. If we represent by 
\(\vartheta\)  the latitude on Earth where the instrument is located, then Eq.~(\ref{eq:echfac}) becomes
\begin{equation}
 \Phi = \frac{4}{\pi} \frac{M}{\hbar} v_{\rm r}^2 (2T)^2 \Omega \sin(\vartheta),
\label{eq:echfac1}
\end{equation}
and consequently we  have
\begin{equation}
 \frac{d P(\Omega)}{d\Omega}\Big|_{\Omega=0}=\frac{2}{\pi}N \eta \frac{M}{\hbar} v_{\rm r}^2 (2T)^2 \sin(\vartheta).
\label{eq:sensiom1}
\end{equation}
If the probability noise is  defined solely by the quantum projection noise, then 
\begin{equation}
 \delta P(\Omega=0)=\sqrt{P(\Omega=0)}=\sqrt{\frac{N}{2}},
\label{eq:sensiom2}
\end{equation}
so we finally find the following expression for the rotation-rate sensitivity:
\begin{equation}
 \delta \Omega =\frac{\pi}{2\eta\sqrt{2N} (M/\hbar) v_{\rm r}^2 (2T)^2 \sin(\vartheta)}.
\label{eq:sensiom3}
\end{equation}

Physically, Eq.~(\ref{eq:sensiom3}) gives us the minimal rotation rate that can be detected 
if the AI is projection-noise limited. To reach the projection-noise-limited 
sensitivity, several technical problems must be overcome. In particular, for a magnetic guide produced with modulated 
currents, the relative stability of the currents supplied to the microwires and the noise level of the phase difference between them must be considered. Both noise sources  induce a fluctuation in 
the separation between the guide and the chip surface.~\cite{reichel} Consequently, the  radial frequency of the guide (i.e., the radial 
confinement energy of 
the guide)  also fluctuates, generating a phase noise that is mapped to the phase of the atomic wave function. As a 
result, the measured inertial phase will acquire a parasitic contribution due to these physical effects. In addition, if the 
guide uses a self-generated offset field,~\cite{clgapra} then the phase fluctuations will also compromise 
the stability of the magnetic guide. Other relevant problems limiting the sensitivity and accuracy of this device, and that  
 are common to cold atom sensors, are detection noise, magnetic-field noise, cloud temperature, and shot-to-shot 
atom number fluctuations (see, for instance, Ref.~\onlinecite{szmuk}).

\begin{figure}[htb]
 \begin{center}
 \includegraphics[width=\columnwidth]{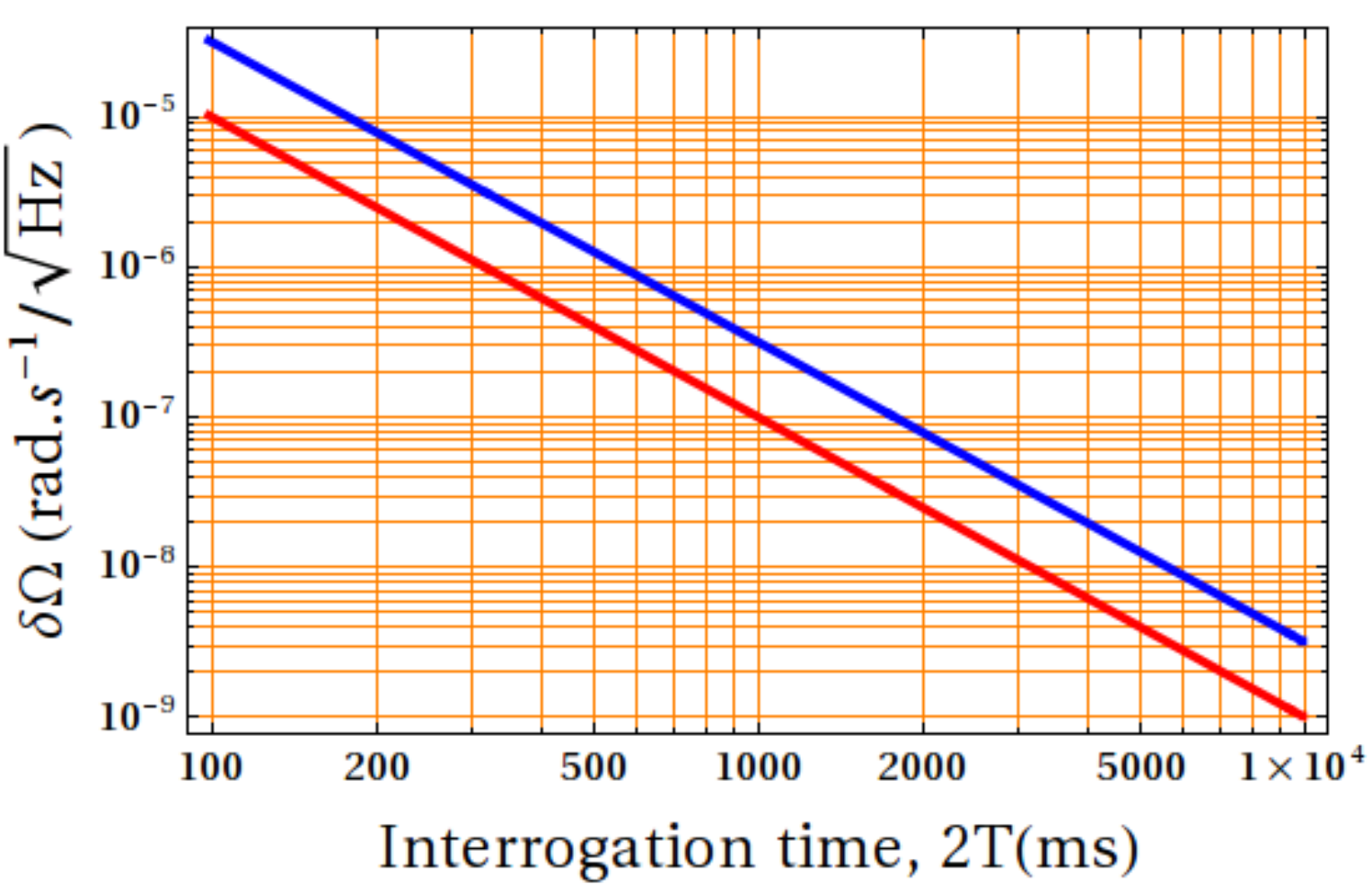}
 \caption{Calculated interferometer rotation-rate sensitivity  vs   interrogation time. The red (blue) line is for  \(N=10^5\ (10^4)\) atoms and perfect visibility (\(\eta=100\%\)).}
 \label{fig:sensiom}
 \end{center}
\end{figure}

As an example,  Fig.~\ref{fig:sensiom} plots the rotation sensitivity given by 
Eq.~(\ref{eq:sensiom3})  versus the interferometer 
interrogation time and  for a fixed launching velocity 2\(v_{\rm r}\), which means that the guide radius depends on the interferometer duration under consideration. 
The result presented in Fig.~\ref{fig:sensiom} indicates that, for instance, using \(10^4\) atoms and an interferometer duration 
 \(2T=3\)~s gives a short-term rotation sensitivity of 
\(3.4\times10^{-8}\)~rad\,s\,\(^{-1}{\rm Hz}^{-1/2}\) or an 
angular random walk,  \(ARW\approx 1.1\times 10^{-4}{} ^\circ\,{{\rm h}}^{-1/2}\). In other words, after 1~h of integration the 
angular standard deviation will be \(1.1\times 10^{-4}{} ^\circ\). Such a rotation-rate sensitivity can be realized by using a 6-mm-radius guide 
 or after 10 round trips in a 600-\(\mu\)m-radius  guide. For defense applications, this ARW is already 
compatible with strategic-grade inertial navigators.

\subsubsection*{Sensitivity function}
This function is defined as
\begin{equation}
 g(t) \equiv 2 \lim_{\delta \phi \to 0} \frac{\delta P}{\delta \phi},
\end{equation}
where \(\delta P\) is the infinitesimal variation of the probability \(P\) of finding the atoms at the AI output port. This 
variation results from an infinitesimal jump of the global phase \(\delta \phi\) due to a perturbation during the 
measurement process. 

\begin{figure}[htb]
 \begin{center}
 \includegraphics[width=\columnwidth]{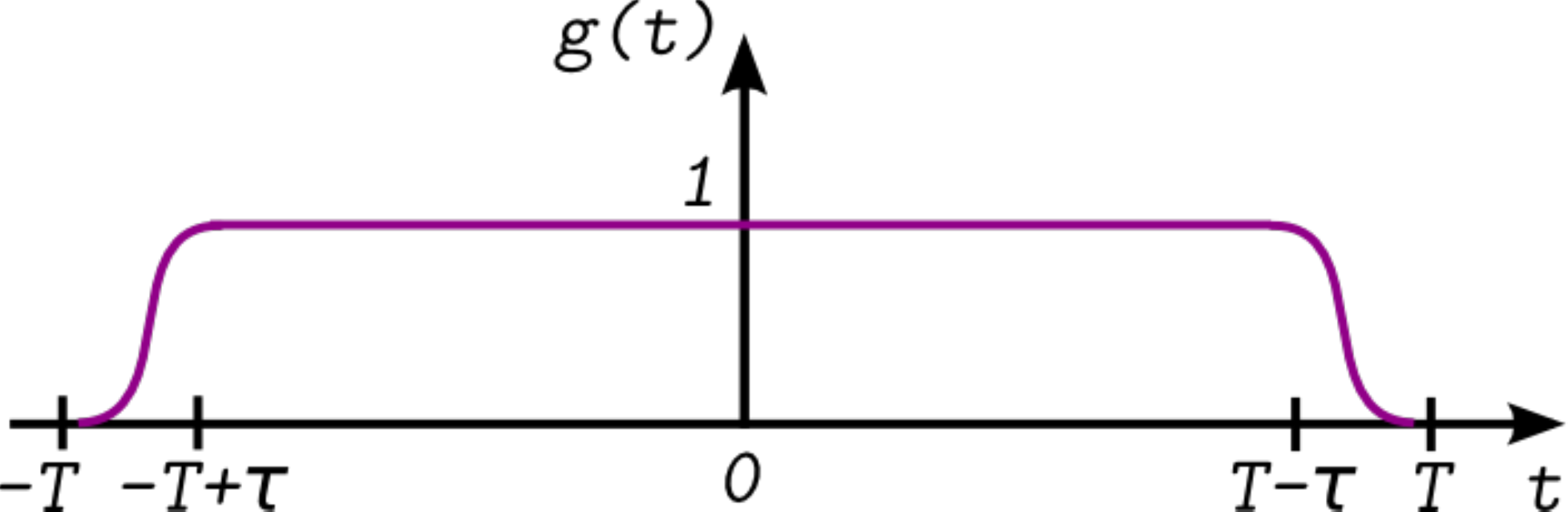}
 \caption{Sensitivity function for the circular guide shown in Fig.~\ref{fig:3w}.}
 \label{fig:sensib}
 \end{center}
\end{figure}

The function \(g(t)\) is a measure of the impact of a given perturbation, which happens 
during the interrogation time \(2T\), on the determination of the atomic phase.~\cite{cheinet} In fact, inside the time window defined 
by \(2T\), the global phase is sensed {\it only} during the time interval defined by the beam-splitter and mirror 
light pulses. Therefore, we expect the sensitivity function to vary during the pulses and to be extremal between 
them. The simplest gyrometer configuration realized with a circular magnetic guide uses two \(\pi/2\) pulses of 
duration \(\tau\), applied at the beginning and at the end of the AI. Therefore, the sensitivity function of this device 
has the form presented in Fig~\ref{fig:sensib}.

\subsubsection*{Transfer function}
To determine the transfer function we  compute the interferometer phase noise 
\(\sigma_{\phi}^{2}\). We denote by \(\ovl{\Delta\phi}\)  the difference of the mean phases during the pulses:
\begin{equation}
 \ovl{\Delta\phi}(\tau)\equiv\ovl{\phi}_{2}(\tau)-\bar{\phi}_{1}(\tau),
\end{equation}
where
\begin{equation}
 \ovl{\phi}_{1}(\tau)=\frac{1}{\tau}\int_{-T}^{-T+\tau}dt\phi(t) ,\quad 
 \ovl{\phi}_{2}(\tau)=\frac{1}{\tau}\int_{T-\tau}^{T}dt\phi(t).
\end{equation}
In this case, the phase noise measured during the pulses is
\begin{equation}
 \sigma_{\phi}^{2}=
 \Big\langle \Big|\frac{1}{\tau}\int_{T-\tau}^{T}dt \phi(t)-\frac{1}{\tau}\int_{-T}^{-T+\tau}dt \phi(t)\Big|^2 \Big\rangle_{T},
\label{eq:signoi}
\end{equation}
where \(\sigma_{\phi}^{2}=\big\langle \big|\ovl{\Delta\phi}(\tau)\big|^{2}\big\rangle_{T}\) and \(\langle \dots \rangle_{T}\) 
is the temporal mean evaluated over a time interval equal to the interrogation time \(2T\). Here, \(\phi(t)\) is the 
instantaneous phase ``seen'' by the interferometer (resulting from rotations, vibrations, laser phase noise, etc.).

\begin{figure}[htb]
 \begin{center}
 \includegraphics[width=\columnwidth]{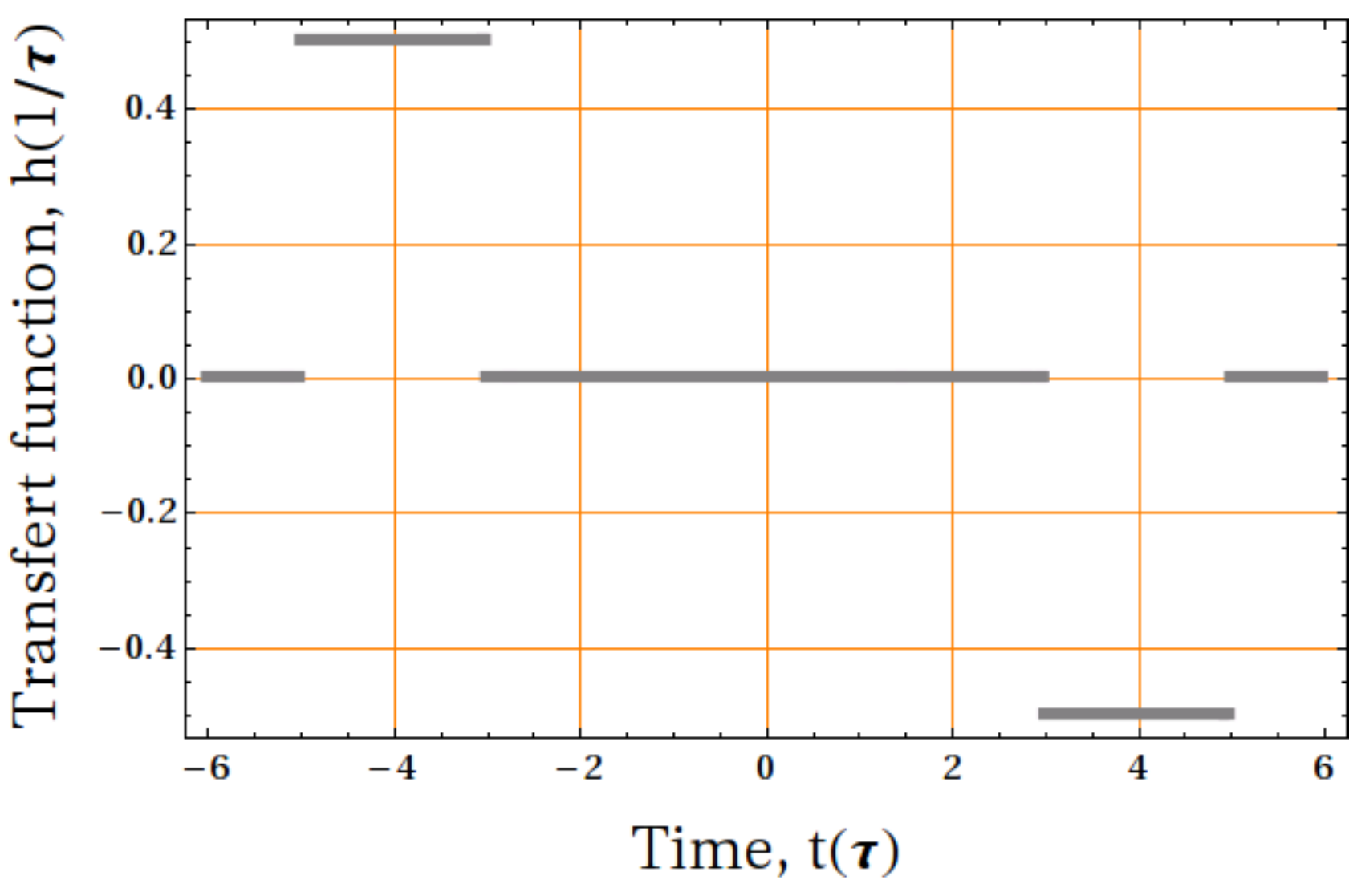}
 \caption{Time-domain transfer function of a two-pulse AI rotation sensor.}
 \label{fig:transf}
 \end{center}
\end{figure}

From Eq.~(\ref{eq:signoi}), the transfer function of this AI in the time domain is \(h(t)\), which is shown in Fig.~\ref{fig:transf} 
and defined as follows:
\begin{equation*}
h(\tau-t)\equiv \left\{
\begin{array}{rl}
0, & |t| < -T\\
1/\tau, & -T \leq t \leq -T+\tau\\
0, & -T+\tau< t < T-\tau\\
-1/\tau, & T-\tau \leq t \leq T.
\end{array} \right.
\end{equation*}
The transfer function \(h(t)\)  expresses the fact that the AI behaves as a bandpass 
filter with characteristic frequencies defined by \(\tau\) and \(T\).

\begin{figure}[htb]
 \begin{center}
 \includegraphics[width=\columnwidth]{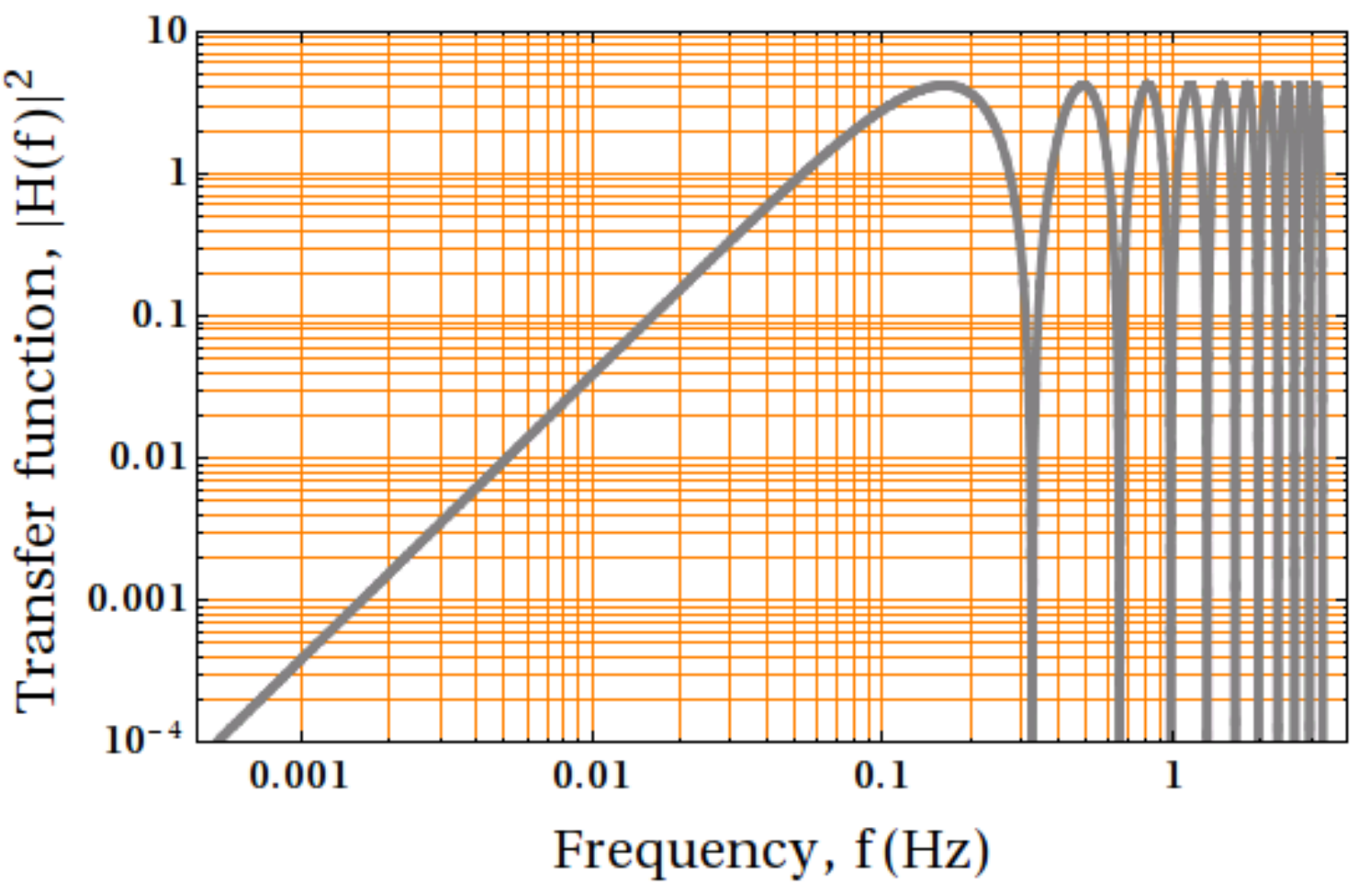}
 \caption{Frequency-domain transfer function of a two-pulse AI. To obtain this plot, we consider a pulse duration \(\tau=20~\mu\)s and 
 an interrogation time    \(2T=4\)~s. The cutoff frequencies are \(f_{HP}=15.9\) kHz 
 and \(f_{LP}=0.1\) Hz.}
 \label{fig:transf1}
 \end{center}
\end{figure}

In the frequency domain, the transfer function \(H(f)\) can be easily computed and is given by the expression
\begin{equation}
 H(f)=-\frac{2\imath}{\pi f \tau}\sin(\pi f \tau)\sin\big[(\pi f(2T-\tau)\big].
\end{equation}
The transfer function is presented in Fig.~\ref{fig:transf1} for experimentally accessible parameter values. The cutoff frequencies are 
given by \(f_{HP}\equiv 1/(\pi \tau)\) and \(f_{LP}\equiv 1/[\pi(2T-\tau)]\). As can be seen in Fig.~\ref{fig:transf1}, an 
ensemble of frequencies exists at which the interferometer is not sensitive to phase noise. These frequencies correspond to multiples of 
the inverse of the pulse duration and the interrogation time.

\subsection{Main sources of noise and systematic effects}
\subsubsection*{Phase noise expressed via the transfer function}
By using the time-domain transfer function,  
Eq. (\ref{eq:signoi}) can be written as
\begin{equation}
 \sigma_{\phi}^{2}=\Big\langle \Big|\int_{-\infty}^{+\infty}dt \phi(t)h(\tau-t)\Big|^2 \Big\rangle_{T}=
 \big\langle \big|\phi(t)\otimes h(t)\big|^2 \big\rangle_{T},
\end{equation}
or, making explicit the temporal mean,
\begin{equation}
 \sigma_{\phi}^{2}=\lim_{T \to \infty} \frac{1}{2T}\int_{-T}^{T}dt \big|\phi(t)\otimes h(t)\big|^2.
\label{eq:signoi1}
\end{equation}

For practical purposes, we  write the phase noise in terms of the interferometer spectral properties, its 
frequency domain transfer function \(H(f)\), and the power spectral density (PSD) of the phase noise, \(S_{\phi}(f)\). By using 
Parseval's theorem and the definition of \(h(t)\),  Eq. (\ref{eq:signoi1}) takes the form
\begin{widetext}
\begin{equation}
 \sigma_{\phi}^{2}=\lim_{T \to \infty} \frac{1}{2T}\int_{-\infty}^{+\infty}dt \big|\phi(t)\otimes h(t)\big|^2 
 \theo^{\rm Parseval}_{\rm theorem}\ \lim_{T \to \infty} \frac{1}{2T}\int_{-\infty}^{+\infty}df \big|\phi_{T}(f)\big|^2 
 \big|H(f)\big|^2,
\end{equation}
\end{widetext}
where \(\phi_{T}(f)\) denotes the Fourier transform of \(\phi(t)\). Therefore, if \(S_{\phi}(f)\) is the PSD of the phase noise, then 
we obtain
\begin{equation}
 \sigma_{\phi}^{2}=\int_{0}^{+\infty}df S_{\phi}(f) \big|H(f)\big|^2.
\label{eq:signoi2}
\end{equation}

\subsubsection*{Sensitivity of AI phase to laser phase noise}
To characterize the noise in the phase accumulated during the 
interferometer interrogation time \(\Phi\), we  use the sensitivity function. Recall that, for a 100\% 
contrast interferometer operating at mid-fringe, the probability \(P\) at the output port is
\begin{equation}
 P=\frac{1}{2}\big[1-\cos(\Phi+\pi/2)\big].
\end{equation}
Consequently, the variation of \(P\) due to an infinitesimal phase jump of the Bragg laser is
\begin{equation}
 \frac{\delta P}{\delta \phi}=\frac{1}{2}\sin(\Phi+\pi/2)\Big|_{\Phi=0}\frac{\delta \Phi}{\delta \phi}=
 \frac{1}{2}\frac{\delta \Phi}{\delta \phi},
\end{equation}
from which the sensitivity function associated with a phase jump becomes
\begin{equation}
 g_{\phi}(t)=\lim_{\delta \phi \to 0}\frac{\delta \Phi}{\delta \phi}.
\end{equation}

Thus, the measured accumulated AI phase during the interrogation time \(2T\) is
\begin{equation}
 \Phi=\int_{-T}^{T}dt g_{\phi}(t)\frac{d}{dt}(\delta\phi),
\end{equation}
and consequently the phase noise after one interferometer cycle is
\begin{equation}
 \sigma_{\Phi}^{2}=\Big\langle \Big|\int_{-T}^{T}dt g_{\phi}(t) \frac{d}{dt}(\delta\phi)\Big|^2 \Big\rangle_{T}.
\end{equation}

By using the definition of the sensitivity function (Fig.~\ref{fig:sensib}) and  considering that this 
function vanishes outside the interval \([-T;T]\), we find
\begin{eqnarray}
 \sigma_{\Phi}^{2} &=& \lim_{T \to \infty} \frac{1}{2T}\int_{-\infty}^{+\infty}dt \big|g_{\phi}(t)\otimes 
\frac{d}{dt}(\delta\phi)\big|^2 \nonumber \\
 &=& \lim_{T \to \infty} \frac{1}{2T}\int_{-\infty}^{+\infty}d\omega \big|G(\omega)\big|^2 
 \big|\omega\ \delta\phi_{T}(\omega)\big|^2,
\end{eqnarray}
and finally
\begin{equation}
 \sigma_{\Phi}^{2}=\int_{0}^{+\infty}d\omega S_{\phi}(\omega)\omega^2 \big|G(\omega)\big|^2.
\label{eq:signoi3}
\end{equation}

Comparing  Eq. (\ref{eq:signoi2}) with Eq. (\ref{eq:signoi3}) shows that the frequency-domain transfer function of the 
AI and the Fourier transform of the sensitivity function \(G(\omega)\) are linked by
\begin{equation}
 \big|H(\omega)\big|^2 = \omega^2 \big| G(\omega)\big|^2.
\end{equation}

In addition, considering Eqs.~(\ref{eq:echfac1}) and (\ref{eq:signoi3}),  the rotation rate measured by 
the AI may be characterized by the  standard deviation
\begin{equation}
 \sigma_{\Omega}=\frac{\pi}{4}\frac{\hbar}{M}\frac{1}{ v_{\rm r}^2 (2T)^2 \sin(\vartheta)}\sigma_{\Phi}.
\label{eq:rotvar}
\end{equation}

\subsubsection*{Sensitivity to vibrations or acceleration noise}
We have previously seen that the AI accumulated phase is
[since \(g_{\phi}(t)=0\ \forall\ t\notin[-T;T]\)\,]
\begin{equation}
 \Phi=\int_{-\infty}^{+\infty}dt g_{\phi}(t)\frac{d}{dt}(\delta\phi).
\end{equation}
In the case of Bragg transitions with counterpropagating laser beams in the \(z\) direction, the phase (wavefront or equiphase 
plane) at the location of the atoms and at  time  \(t\) is \(\phi(t)=2 k z(t)\). Thus,
\begin{equation}
 \Phi=2k\int_{-\infty}^{+\infty}dt g_{\phi}(t)\frac{d}{dt}(\delta z).
\label{eq:phiac}
\end{equation}
Computing the integral in Eq. (\ref{eq:phiac}) gives
\begin{equation}
 \Phi=\int_{-\infty}^{+\infty}dt\Bigg[-2k\int_{-\infty}^{t}dt' g_{\phi}(t')\Bigg]\delta\Big(\frac{d^2 z}{dt^2}\Big),
\label{eq:phiacc}
\end{equation}
which leads to the following definition of the sensitivity function of the AI to accelerations:
\begin{equation}
 g_{a}(t) \equiv -2k \int_{-\infty}^{t}dt' g_{\phi}(t'),
\end{equation}
so that the interferometer phase due to an acceleration can be written in the compact form
\begin{equation}
 \Phi_{a}=\int_{-\infty}^{+\infty}dt g_{a}(t)\delta a(t).
\label{eq:phiacc1}
\end{equation}

From the physics point of view, Eq.~(\ref{eq:phiacc1}) is the sum of all the infinitesimal acceleration variations 
\(\delta a(t)\) experienced by the device. These contributions are taken at  time  \(t\) and weighted by 
the sensitivity function to accelerations; the latter is also evaluated at the same time instant. The noise in the measured phase 
at the output of an AI experiencing vibrations is therefore
\begin{equation}
 \sigma_{\Phi_{a}}^{2}=\Big\langle \Big|\int_{-T}^{T}dt g_{a}(t) \delta a(t)\Big|^2 \Big\rangle_{T}.
\label{eq:phiacc2}
\end{equation}

Considering the relationship between the acceleration and the phase \(\phi(t)\), we can rewrite Eq.~(\ref{eq:phiacc2}) in the 
form
\begin{equation}
 \sigma_{\Phi_{a}}^{2}=\Big\langle \int_{-\infty}^{+\infty}dt \Big|\frac{1}{2k}g_{a}(\tau-t) 
 \frac{d^2}{dt^2}(\delta\phi) \Big|^2 \Big\rangle_{T}.
\end{equation}
Next, by using the definition for the temporal mean and for the convolution product, we get
\begin{eqnarray}
 \sigma_{\Phi_{a}}^{2} &=& \lim_{T \to \infty} \frac{1}{2T}\int_{-\infty}^{+\infty}dt 
 \Big|\frac{1}{2k} g_{a}(t)\otimes \frac{d^2}{dt^2}(\delta\phi)\Big|^2.
\end{eqnarray}
As before, we can use Parseval's theorem  to find the final expression for the vibration phase noise. It is given 
by
\begin{equation}
 \sigma_{\Phi_{a}}^{2}=\int_{0}^{+\infty}d\omega \frac{\omega^4}{(2k)^2}S_{\phi}(\omega) \big|G_{a}(\omega)\big|^2.
\label{eq:phiacc3}
\end{equation}

Equation~(\ref{eq:phiacc3}) has a simple physical meaning: \(
{(\sigma_{\Phi_{a}}^{2})^{1/2}}\) is the  rms radians added 
by vibrations or acceleration noise to the atomic phase one would like to measure with the AI. 

Again, by comparing Eq.~(\ref{eq:phiacc3}) with Eq.~(\ref{eq:signoi3}), we can derive the following practical expressions linking 
the spectral properties of the acceleration noise with those of the phase noise. In fact, isolating 
\(\big|G_{a}(\omega)\big|^2\) in Eq. (\ref{eq:phiacc3}), we find
\begin{eqnarray}
 \big|G_{a}(\omega)\big|^2 &=& \frac{(2k)^2}{\omega^2} \big|G_(\omega)\big|^2,\\
 S_{a}(\omega) &=& \frac{\omega^4}{(2k)^2}S_{\phi}(\omega),\\
 \big|H_{a}(\omega)\big|^2 &=& \frac{(2k)^2}{\omega^4} \big|H(\omega)\big|^2.
\end{eqnarray}
Thus, if we independently measure the power spectral density of the interferometer vibrations \(S_{a}(\omega)\), 
then we get [because \(H(\omega)\) can always be computed]
\begin{equation}
 \sigma_{\Phi_{a}}^{2}=\int_{0}^{+\infty}d\omega \frac{(2k)^2}{\omega^4} S_{a}(\omega) \big|H(\omega)\big|^2
\label{eq:phiacc4}
\end{equation}
for the  rms vibration phase in radians.

\subsubsection*{Sensitivity to rotation noise}
Here we  refer to the rotation noise caused by the fluctuations of the 
interferometer 
rotation sensing axis i.e., the axis perpendicular to the oriented area of the interferometer. The starting point in this calculation is the scale factor ~(\ref{eq:echfac}). Writing this 
equation as
\begin{equation}
 \phi(t)=4 k R \theta(t)
\end{equation}
with \(\theta(t)=\Omega t\), we find the following expression for the accumulated phase at the end of the interferometer cycle: 
\begin{equation}
 \Phi=4 k R\int_{-\infty}^{+\infty}dt g_{\phi}(t)\frac{d}{dt}(\delta \theta).
\label{eq:phirot1}
\end{equation}
After defining the  function of sensitivity to rotations as \(g_{\Omega}(t) \equiv 4 k R g_{\phi}(t)\), Eq. (\ref{eq:phirot1}) becomes
\begin{equation}
 \Phi=\int_{-\infty}^{+\infty}dt g_{\Omega}(t) \delta \Omega(t).
\label{eq:phirot2}
\end{equation}

In analogy with the derivation of Eq. (\ref{eq:phiacc3}), we  show that the measured rotation noise is given by
\begin{equation}
 \sigma_{\Phi_{\Omega}}^{2}=\int_{0}^{+\infty}d\omega \frac{\omega^2}{(4k R)^2}S_{\phi}(\omega) 
 \big|G_{\Omega}(\omega)\big|^2,
\label{eq:phirot3}
\end{equation}
where \(G_{\Omega}(\omega)\) is the Fourier transform of the rotation-sensitivity function. In the present case, the 
useful relationships between the spectral properties of the phase noise and the rotation noise are
\begin{eqnarray}
 \big|G_{\Omega}(\omega)\big|^2 &=& (4 k R)^2 \big|G_(\omega)\big|^2,\\
 S_{\Omega}(\omega) &=& \frac{\omega^2}{(4 k R)^2} S_{\phi}(\omega),\\
 \big|H_{\Omega}(\omega)\big|^2 &=& \frac{(4 k R)^2}{\omega^2} \big|H(\omega)\big|^2.
\end{eqnarray}

Once again, if one determines the PSD of the rotation noise \(S_{\Omega}(\omega)\), 
then, from the equation
\begin{equation}
 \sigma_{\Phi_{\Omega}}^{2}=\int_{0}^{+\infty}d\omega \frac{(4 k R)^2}{\omega^2} S_{\Omega}(\omega) \big|H(\omega)\big|^2,
\label{eq:phirot4}
\end{equation}
we  find  \(
{(\sigma_{\Phi_{\Omega}}^{2})^{1/2}}\)    rms radians of rotation noise contributing to the measured phase 
signal of the interferometer.

\subsubsection*{Stability}
\begin{figure}[htb]
 \begin{center}
 \includegraphics[width=1.05\columnwidth]{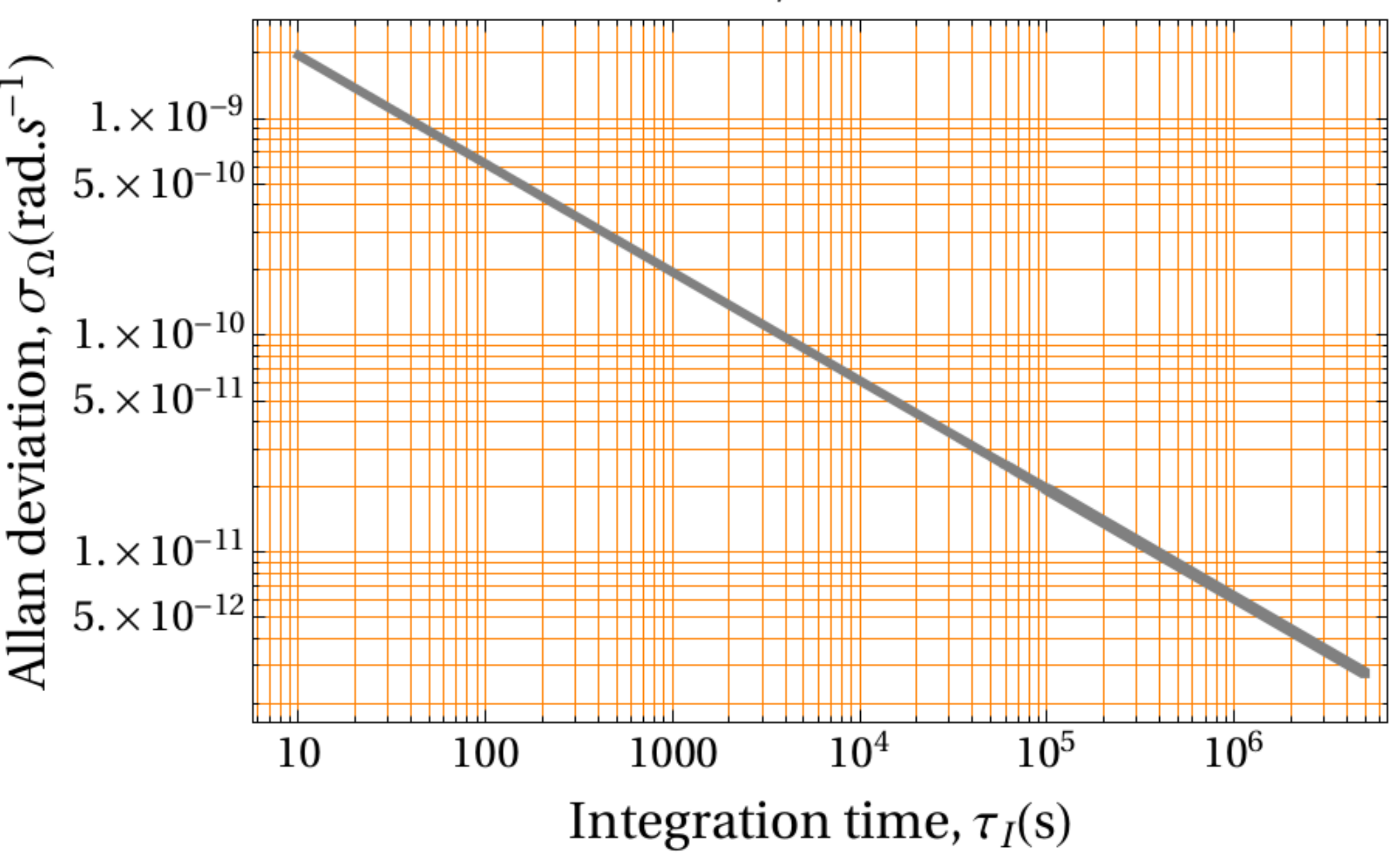}
 \caption{Computed rotation rate in the form of the Allan standard deviation for a projection-noise-limited 
 AI using a circular waveguide.}
 \label{fig:allanrot}
 \end{center}
\end{figure}

As  is well known, the most informative quantity 
about the stability of a sensor 
(or instrument in general) is the Allan variance.~\cite{allandev} By using equations (\ref{eq:rotvar}) and 
(\ref{eq:phirot4}),  the Allan variance for the measured rotation rate with this AI may be shown to be
\begin{eqnarray}
\label{eq:allanrot}
 \sigma_\Omega^2(\tau_I) &=& \Big[ \frac{\pi}{4}\frac{\hbar}{M}\frac{1}{ v_{\rm r}^2 (2T)^2 \sin(\vartheta)}\Big]^2 \\
 &&\times\frac{4\pi}{\tau_I}\sum_{m=0}^{\infty}\frac{(4 k R)^2}{[2\pi m/(2T)]^2} \big|H(m/T)\big|^2 
 S_{\Omega}(m/T) ,\nonumber
\end{eqnarray}
where \(\tau_{I}\) is the integration time. To obtain an order of magnitude of this quantity, let us consider a 
projection-noise-limited AI with \(2T=10\)~s of interrogation time.~\cite{note0} If 
\(10^5\) atoms are launched at \(2v_{\rm r}\) by a \(\pi/2\) pulse with a 
duration of \(\tau=20~\mu\)s, then the  Allan standard deviation is 
\(1.9\times10^{-9}~{\rm rad\, s^{-1}}/\sqrt{\tau_I({\rm s})}\).  Figure~\ref{fig:allanrot} shows the computed Allan standard deviation 
 for this particular example.~\cite{note1} 

Note that, after 
12 months of integration, the interferometer reaches a stability of 
\(3.5\times10^{-13}~{\rm rad\, s^{-1}}\), which is 
theoretically compatible with applications in geophysics and the realization of tests in fundamental physics, such as the 
observation of the geodetic effect.
In fact, in the 1960s, using general relativity, Leonard Schiff predicted that a free-falling inertial frame in a polar orbit 
around a rotating gravitational source  experiences two orthogonal rotations with respect to the fixed inertial frame of 
the Universe.~\cite{schiff,schiff1} These two phenomena, called the geodetic and 
Lense-Thirring effects, are characterized by 
rotation rates of 6.6 arcseconds/year and 33 milliarcseconds/year, respectively (1 milliarcsecond = 
\(4.848\times10^{-9}\)~rad\,s\(^{-1}\)), for an orbit located 642 km  from the Earth.

Recently, in 2011, the Gravity Probe B (GP-B) experiment developed by Stanford University 
and NASA confirmed these predictions with 
a precision of 1\% by using a satellite. The science (data recording) phase of this mission lasted 353 
days.~\cite{GPB,GPB1} If, in an analogous way, we would like to measure, for instance, the geodetic effect by using cold guided atoms on an atom 
chip with a precision of 5\% in 1 year,  then we would need a gyrometer with a stability of about  
\(5.2\times10^{-14}\)~rad\,s\(^{-1}\). From the result presented in Fig.~\ref{fig:allanrot} and using the parameter values 
stated above, such a measurement would require 47 years!

\begin{figure}[htb]
 \begin{center}
 \includegraphics[width=\columnwidth]{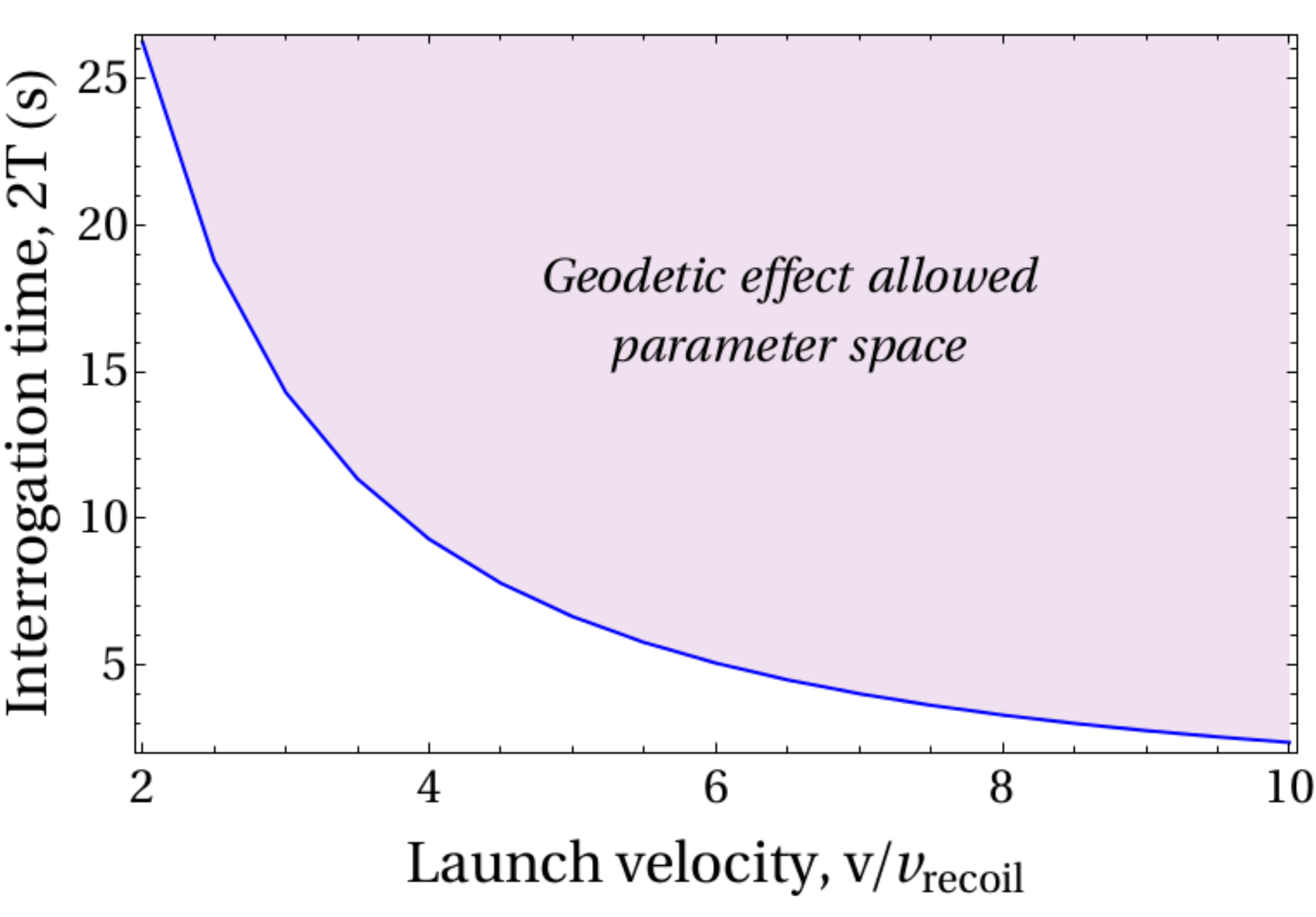}
 \caption{Experimental limits of the launching speed and the AI interrogation time compatible with the observation of 
 the geodetic effect with 5\% precision in 12 months of integration.}
 \label{fig:diagph}
 \end{center}
\end{figure}

However,  this problem should be tackled in a different way. The meaningful question is whether we can  design a compact 
atom-chip-based gyrometer by using cold guided atoms to reliably measure  the geodetic effect from a satellite 
orbiting at 642 km above  Earth. Part of the answer to this question is presented in Fig.~\ref{fig:diagph}, which 
shows the  \(v\text{-}2T\) diagram (i.e., interrogation time versus launching speed) with a 5\% precision  for 12 months of integration 
time.
We  learn from Fig.~\ref{fig:diagph} that, if the atoms are launched with, say, an initial velocity of 
4\(v_{\rm r}\), then we  need a minimum interrogation time \(2T=9\)~s  to achieve  5\% precision in 
the measurement. For this particular case, we would need a 37-mm-radius circular magnetic guide. 

\begin{figure}[htb]
 \begin{center}
 \includegraphics[width=\columnwidth]{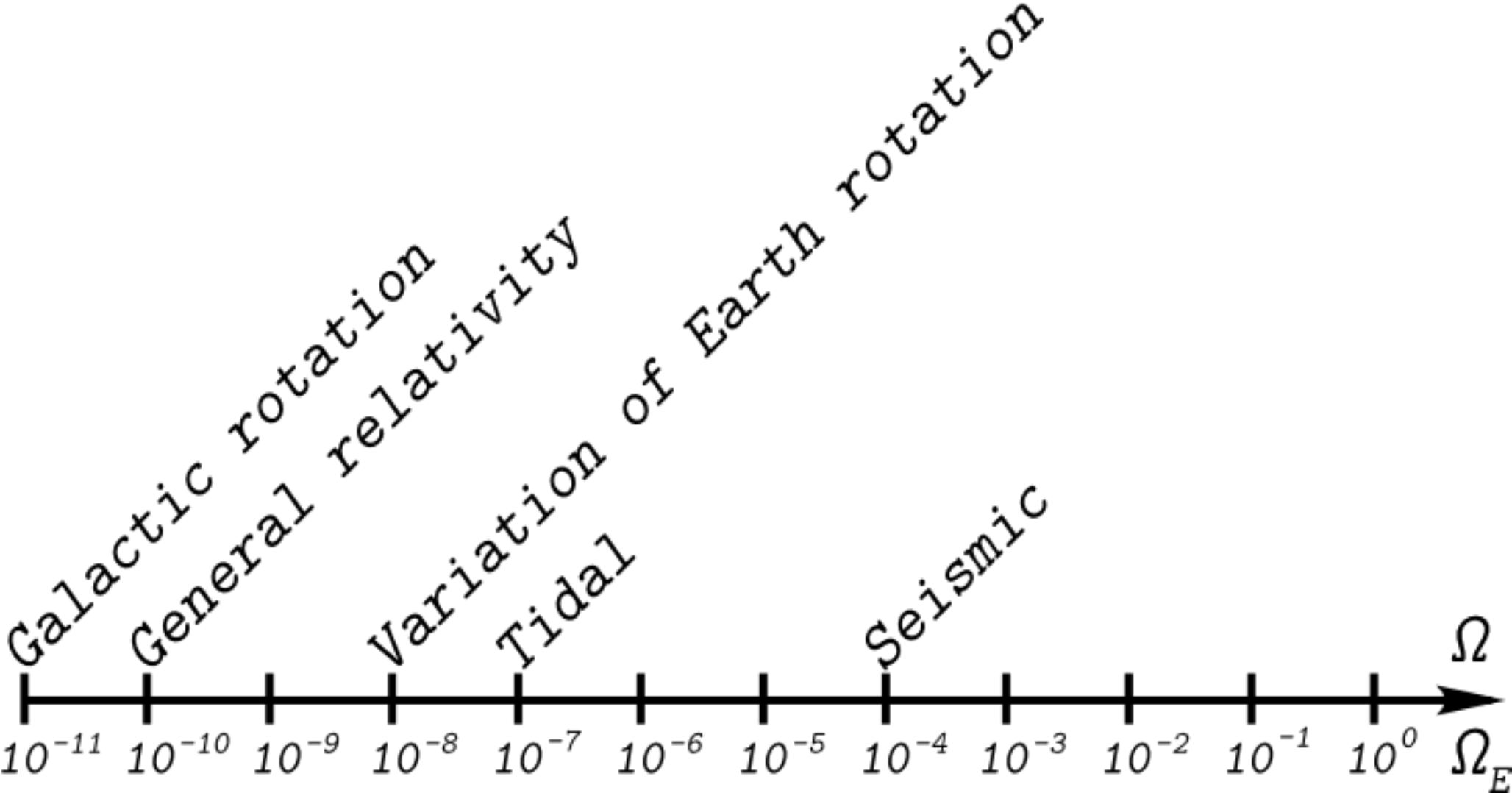}
 \caption{Order of magnitude of  rotation rates characteristic of different phenomena in geophysics and fundamental 
 physics. These rotation rates are scaled to the rotation rate of Earth: \(\Omega_{\rm E}=7.29\times10^{-5}\)~rad\,s\(^{-1}\).}
 \label{fig:geophys}
 \end{center}
\end{figure}

Suppose now that we can develop an atom chip gyrometer compatible with the CubeSat 
technology.~\cite{cubesat} In such a scenario, we could foresee, for example, the simultaneous measurement of the geodetic effect 
using two cubesats orbiting at different distances from  Earth. The more distant AI could  provide a reference 
measurement in the common frame of the two satellites, and a differential signal could  provide  better precision 
to demonstrate this effect of general relativity. Moreover, if a multi-axis atom-chip-based inertial sensor is developed, 
then it could be used to realize a drag-free satellite configuration for this scenario. Finally, to have 
an idea of the potential applications accessible with this AI configuration,  Fig.~\ref{fig:geophys} shows the order of 
magnitude of the rotation rate associated with different fundamental physics phenomena.

An important point to have in mind in the design of a cold atom sensor is the scale of the instrument in terms 
of volume, power consumption, and weight. For an atom-chip-based sensor, the core of the physics package 
(i.e., the chip) is a few square centimeters, its power consumption can be kept below 1 W, and, considering the holder, it can weigh up to a 
few hundred grams. The next important element in the physics package is the vacuum system. By using a glass cell, its 
volume can be as small as 10~cm\(^3\). However, a physics package with such a small vacuum chamber would certainly require a 
second chamber to realize a double-MOT configuration to avoid  reducing the lifetime of the cold samples. In fact, the vacuum system plays a key role in determining the dead time between measurements.~\cite{Dugrain,jmm} For instance,  assume that the interferometer  uses molasses cooled atoms. If we use a single vacuum chamber, then the expected dead 
time is on the order of 5~s. If we use a double-chamber vacuum system (2DMOT + 3DMOT), then the dead time can be reduced 
to less than 1~s. In both cases, if we use evaporatively cooled atoms, then we need to add, typically, at least 3~s to the 
previous values. From the reported results (see, for instance, Ref.~\onlinecite{grosse}), one can expect that a physics package of 1 L 
and 1 kg is feasible for this type of sensor.

Another key element of the sensor is the laser source. Today, low-phase-noise transportable laser sources are 
commercially available.~\cite{muquans} To best exploit the compactness of  atom chips, an important technological 
effort is required to integrate such laser sources onto an atom chip substrate.

Two other important sensor requirements for inertial navigation and 
guidance are the bandwidth and the dynamic range. High-data-rate sensors are demonstrated in Refs. ~\onlinecite{rakholia,savoie}, and a sensor 
with relatively large dynamic range  was demonstrated in Ref.~\onlinecite{barret}. These experimental realizations are very encouraging 
because, given the time required to cool the atoms, inertial sensors using  atom interferometry technology have in general a 
very small bandwidth, typically below 1 Hz. This is a clear drawback if the sensor is expected, for instance, to provide 
guidance to a 
carrier. In addition, the performance (sensitivity and stability) of a cold atom inertial sensor is dramatically affected 
by vibrations. Therefore, one might not expect cold atom sensors to supplant the conventional inertial sensors that are commercially available  today (i.e., ring laser gyros, fiber-optic gyros, MEMS accelerometers, and gyroscopes). However, contrary to cold atom inertial 
sensors, conventional sensors lack the requisite stability for long-term navigation,  are not absolute,  require 
calibration, and 
usually need an external reference signal that is susceptible to jamming. Therefore, as an optimal navigation solution one can 
foresee the hybridization of cold atom and conventional inertial sensors.~\cite{lautier}\\

\section{Outlook and conclusion}
This review presents the state-of-the-art results in atom interferometry that are relevant for inertial 
navigation applications. These results were obtained by using laboratory instruments that, given their volume, are not yet compatible 
with mobile 
applications. However, they are undoubtedly relevant to fundamental studies and to define the 
ultimate performance that can be realized with the compact portable sensors. We also discuss a representative set 
of portable and compact atom interferometers that have demonstrated an inertial sensitivity. In particular, for 
rotation sensing, examples are given for two classes of interferometers: free-falling atoms and guided atom 
interferometers. The latter is considered in detail when using an atom chip as the sensor platform. The enabling technologies and a case study of a sensor design for inertial navigation 
applications are also presented. Notably, from the computed sensitivity in 
Fig.~\ref{fig:sensiom}, the expected angular 
random walk [$1.1\times 10^{-4}$ $^\circ{\rm h}^{-1/2}$]
 after 1~h of integration suggests that a sensor based on the  design under consideration 
would be compliant with strategic-grade inertial navigators.

Projection-noise-limited atom interferometric inertial sensors have now been demonstrated in the free-falling 
configuration. When dealing with atom chips with relatively low atom number, this result suggests that a possible 
solution to reach the desired performance is to implement quantum metrology protocols by using squeezed 
and entangled
states.~\cite{eckert,caves,riedel,reinhard,leo,koschorreck} In this case, the device sensitivity would scale as \(\Delta\Phi_{\rm min} \approx \xi(N)/\sqrt{N}\), 
where \(\xi(N) < 1\) defines the degree of atomic noise squeezing. In the context of atom interferometry, 
these states have been generated in optical dipole traps,~\cite{appel} optical cavities in the QED regime,~\cite{smith} and in 
optical lattices.~\cite{gross} 

\section{Acknowledgments}
I would like to thank the members of the atom interferometry and inertial sensors team at SYRTE.
This work was funded by the D\'el\'egation G\'en\'erale de l'Armement (DGA)
through the ANR ASTRIDE program (Contracts ANR-13-ASTR-0031-01, ANR-18-ASMA-0007-01), the Institut Francilien de Recherche 
sur les Atomes Froids (IFRAF), and the Emergence-UPMC program (Contract A1-MC-JC-2011/220).

 

\begin{thebibliography}{99}
\bibitem{note2} For example, the power dissipated on a gold microwire of \(3\times3\) \(\mu\)m\(^2\) cross section fed with 1 A 
can be as high as 10 W.
\bibitem{note1} We have assumed a white noise PSD, \(S_{\Omega}(m/T)\) in Eq. (\ref{eq:allanrot}) equals to 
\((\delta \Omega)^2/{\rm Hz}\) - cf. Eq. (\ref{eq:sensiom3}).
\end{thebibliography}
\end{document}